\documentclass[amsthm,preprint,times]{elsart}
\pdfoutput=1

\usepackage[utf8]{inputenc}
\usepackage[english]{babel} 
\usepackage[T1]{fontenc}
\usepackage{epsfig}
\usepackage{amsmath, amssymb, amsbsy}
\usepackage{xspace}
\usepackage{color}
\usepackage{url}
\usepackage{booktabs}
\usepackage[colorlinks=true,citecolor=blue,linkcolor=blue]{hyperref}
\usepackage[linesnumbered,ruled,vlined,titlenumbered]{algorithm2e}
\usepackage{enumitem}

\newtheorem{theorem}{Theorem}
\newtheorem{lemma}[theorem]{Lemma}
\newtheorem{corollary}[theorem]{Corollary}

\newtheorem{definition}[theorem]{Definition}

\newtheorem{remark}[theorem]{Remark}

\usepackage{varioref}
\usepackage{fancyref}

\makeatletter
\def\mkfancyprefix#1#2{%
\expandafter\def\csname fancyref#1labelprefix\endcsname{#1}%
\begingroup\def\x{\endgroup\frefformat{plain}}%
    \expandafter\x\csname fancyref#1labelprefix\endcsname
    {\MakeLowercase{#2}\fancyrefdefaultspacing##1}%
\begingroup\def\x{\endgroup\Frefformat{plain}}%
    \expandafter\x\csname fancyref#1labelprefix\endcsname
    {#2\fancyrefdefaultspacing##1}%
\begingroup\def\x{\endgroup\frefformat{vario}}%
    \expandafter\x\csname fancyref#1labelprefix\endcsname
    {\MakeLowercase{#2}\fancyrefdefaultspacing##1##3}%
\begingroup\def\x{\endgroup\Frefformat{vario}}%
    \expandafter\x\csname fancyref#1labelprefix\endcsname
    {#2\fancyrefdefaultspacing##1##3}%
}
\makeatother
\fancyrefchangeprefix{\fancyrefeqlabelprefix}{eqn}
\mkfancyprefix{sec}{Section}
\mkfancyprefix{subsec}{Section}
\mkfancyprefix{tbl}{Table}
\mkfancyprefix{thm}{Theorem}
\mkfancyprefix{lem}{Lemma}
\mkfancyprefix{cor}{Corollary}
\mkfancyprefix{prop}{Proposition}
\mkfancyprefix{prob}{Problem}
\mkfancyprefix{alg}{Algorithm}
\mkfancyprefix{inv}{Invariant}
\mkfancyprefix{ex}{Example}
\mkfancyprefix{line}{Line}
\mkfancyprefix{def}{Definition}
\mkfancyprefix{itm}{Item}
\newcommand{\cref}[1]{\Fref{#1}}

\makeatletter
\newcommand{\removelatexerror}{\let\@latex@error\@gobble}
\makeatother

\newcommand{\printalgoIEEE}[1]
{{\centering
\scalebox{0.97}{
\removelatexerror
\begin{tabular}{p{\textwidth}}
\begin{algorithm}[H]
 #1
\end{algorithm}
\end{tabular}
}
}
}

\DeclareMathOperator{\rank}{rank}

\DeclareUnicodeCharacter{0109}{\koppa}

\def\ve#1{{\mathchoice{\mbox{\boldmath$\displaystyle #1$}}%
              {\mbox{\boldmath$\textstyle #1$}}%
              {\mbox{\boldmath$\scriptstyle #1$}}%
              {\mbox{\boldmath$\scriptscriptstyle #1$}}}}

\newcommand{\myset}[1]{\mathcal{#1}}
\newcommand{\intervallincl}[2]{\ensuremath{[#1,#2]}}

\renewcommand{\vec}[1]{\ensuremath{\mathbf{#1}}}
\newcommand{\Mat}[1]{\ensuremath{{#1}}}

\newcommand{\vecelementsArb}[2]{\ensuremath{\big[\begin{matrix}#1_1 & #1_2 & \dots & #1_{#2}\end{matrix}\big]}}

\newcommand{\qreciproc}[1]{\overline{#1}}
\newcommand{\LEEAOutputRx}{\ensuremath{r_\mathrm{out}}}
\newcommand{\LEEAOutputUx}{\ensuremath{u_\mathrm{out}}}
\newcommand{\LEEAOutputU}{\ensuremath{u_\mathrm{out}}}
\newcommand{\LEEAOutputVx}{\ensuremath{v_\mathrm{out}}}

\newcommand{\npluskhalf}{\left\lfloor {(n+k)}/{2}\right\rfloor}
\newcommand{\numbRowErasures}{\varrho}
\newcommand{\numbColErasures}{\gamma}
\newcommand{\Basis}{\ensuremath{\myset{B}}}
\newcommand{\Dualbasis}{\myset{B}^{\perp}}
\newcommand{\Normbasis}{\myset{B}_N}

\newcommand{\Normelement}{\beta}

\newcommand{\mycode}[1]{\ensuremath{\mathcal{#1}}}
\newcommand{\Gab}[1]{\ensuremath{\mycode{G}[#1]}}

\newcommand{\B}{\mathcal{B}}
\newcommand{\K}{K}
\newcommand{\C}{\ve{C}}
\newcommand{\A}{\ve{A}}
\newcommand{\T}{\ve{T}}
\newcommand{\F}{\mathbb{F}}

\newcommand{\N}{\mathbb{N}}
\newcommand{\Fq}{\mathbb{F}_q}
\newcommand{\Fqm}{\mathbb{F}_{q^m}}
\newcommand{\Fqs}{\mathbb{F}_{q^s}}
\newcommand{\qdeg}{\mathrm{deg}_q}

\newcommand{\LH}[1]{\langle #1 \rangle}
\newcommand{\Lset}{\mathcal{L}_{q^m}}
\newcommand{\BigO}[1]{\mathcal{O}\left(#1\right)}
\newcommand{\BigOtext}[1]{\mathcal{O}(#1)}

\newcommand{\BigOmega}[1]{\Omega\left(#1\right)}
\newcommand{\BigTheta}[1]{\Theta\left(#1\right)}
\newcommand{\BigOT}[1]{\mathcal{O}^\sim\left(#1\right)}
\newcommand{\BigOTtext}[1]{\mathcal{O}^\sim(#1)}

\newcommand{\OMul}[1]{\mathcal{M}_{q^m}\left(#1\right)}

\newcommand{\OMulSkew}[1]{\bar{\mathcal{M}}_{q^m}\left(#1\right)}
\newcommand{\ODiv}[1]{\mathcal{D}_{q^m}\left(#1\right)}

\newcommand{\OQT}[1]{\mathcal{QT}_{q^m}\left(#1\right)}

\newcommand{\OMSP}[1]{\mathcal{MSP}_{q^m}\left(#1\right)}
\newcommand{\OMPE}[1]{\mathcal{MPE}_{q^m}\left(#1\right)}
\newcommand{\OIP}[1]{\mathcal{I}_{q^m}\left(#1\right)}

\newcommand{\Lsetmaxs}{\Lset^{\leq s}}

\newcommand{\Lsetsmallerm}{\mathcal{L}_{q^m}^{< m}}

\newcommand{\qtr}[1]{\hat{#1}}
\newcommand{\mul}{\cdot}
\newcommand{\ev}{\mathrm{ev}}

\newcommand{\sstar}{s^\ast}

\newcommand{\Lsmallers}{\Lset^{< s}}
\newcommand{\Lsmallerk}{\Lset^{< k}}

\newcommand{\dR}{\mathrm{d_R}}
\newcommand{\rk}{\mathrm{rk}}

\newcommand{\MSP}[1]{\mathcal{M}_{#1}}
\newcommand{\U}{\mathcal{U}}
\renewcommand{\qed}{\hfill\square}

\newcommand{\AMSP}[1]{\mathrm{MSP}\left( #1 \right)}
\newcommand{\AMPE}[2]{\mathrm{MPE}\left( #1,#2 \right)}
\newcommand{\AIP}[1]{\mathrm{IP}\left( #1 \right)}

\newcommand{\ADIV}[1]{\mathrm{RightDiv}\left( #1 \right)}
\newcommand{\ARDIV}[1]{\ADIV{#1}}
\newcommand{\ALDIV}[1]{\mathrm{LeftDiv}\left( #1 \right)}
\newcommand{\ALEEA}[1]{\mathrm{HalfLEEA}\left( #1 \right)}

\newcommand{\RSpace}[1]{\ker(#1)}
\newcommand{\quo}{\chi}
\newcommand{\remainder}{\varrho}
\newcommand{\qL}{\quo_\mathrm{L}}
\newcommand{\qR}{\quo_\mathrm{R}}
\newcommand{\rL}{\remainder_\mathrm{L}}
\newcommand{\rR}{\remainder_\mathrm{R}}
\newcommand{\gauss}[1]{\lfloor #1 \rfloor}

\newcommand{\IP}[1]{\mathcal{I}_{#1}}
\newcommand{\autom}{\sigma}
\newcommand{\automi}[1]{\sigma^{#1}}

\newcommand{\Aset}{\Fqm[x;\autom]} 
\newcommand{\Asetmaxs}{\Aset_{\leq s}}

\newcommand{\amul}{\cdot}
\newcommand{\supp}{\mathrm{supp}}

\newcommand{\NN}{\mathbb{N}}

\newcommand{\polymulexponent}{{\min\left\{\frac{\omega+1}{2},1.635 \right\}}}

\newcommand{\End}{\mathrm{End}}

\renewcommand{\r}{\ve{r}}
\renewcommand{\e}{\ve{e}}
\renewcommand{\c}{\ve{c}}
\definecolor{darkred}{rgb}{0.7,0,0}

\newcommand{\mulmodm}{\cdot_{\, \mathrm{mod} \, (x^{[m]}-x)}}

\renewcommand{\bar}{\overline}
\renewcommand{\tilde}{\widetilde}

\definecolor{myblue}{rgb}{0,0,0.7}
\definecolor{mygreen}{rgb}{0,0.7,0}
\definecolor{myred}{rgb}{0.7,0,0}

\newboolean{cleanversion}
\setboolean{cleanversion}{false}
\ifcleanversion
\newcommand{\todo}[1]{}
\newcommand{\reviewerone}[1]{}
\newcommand{\reviewertwo}[1]{}
\newcommand{\pu}[1]{}
\newcommand{\aw}[1]{}

\newcommand{\old}[1]{}
\else
\newcommand{\todo}[1]{{\leavevmode\color{red}[{\it #1}]}\xspace}
\newcommand{\pu}[1]{{\leavevmode\color{mygreen}[{\bf pu:} #1]}}
\newcommand{\aw}[1]{{\leavevmode\color{mygreen}[{\bf aw:} #1]}}
\newcommand{\old}[1]{{\leavevmode\color{myred}[{\bf old:} #1]}}
\fi

\usepackage[mathtime]{yjsco}
\def\qed{\relax\ifmmode\hfill \Box\else\unskip\nobreak\hfill $\Box$\fi}
\usepackage{natbib}

\begin{document}

\begin{frontmatter}

\title{Fast Operations on Linearized Polynomials and their Applications in Coding Theory} 
\author{Sven Puchinger}
\address{Institute of Communications Engineering, Ulm University, Ulm, Germany}
\ead{sven.puchinger@uni-ulm.de}
\author{Antonia Wachter-Zeh}
\address{Institute for Communications Engineering, Technical University of Munich, Germany}
\ead{antonia.wachter-zeh@tum.de}

\begin{abstract}
This paper considers fast algorithms for operations on linearized polynomials.
We propose a new multiplication algorithm for skew polynomials (a generalization of linearized polynomials) which has sub-quadratic complexity in the polynomial degree $s$, independent of the underlying field extension degree~$m$.
We show that our multiplication algorithm is faster than all known ones when $s \leq m$.
Using a result by \citet{caruso2017new}, this immediately implies a sub-quadratic division algorithm for linearized polynomials for arbitrary polynomial degree $s$.
Also, we propose algorithms with sub-quadratic complexity for the $q$-transform, multi-point evaluation, computing minimal subspace polynomials, and interpolation, whose implementations were at least quadratic before.
Using the new fast algorithm for the $q$-transform, we show how matrix multiplication over a finite field can be implemented by multiplying linearized polynomials of degrees at most $s=m$ if an elliptic normal basis of extension degree $m$ exists, providing a lower bound on the cost of the latter problem.
Finally, it is shown how the new fast operations on linearized polynomials lead to the first error and erasure decoding algorithm for Gabidulin codes with sub-quadratic complexity.
\end{abstract}

\begin{keyword}
linearized polynomials \sep skew polynomials \sep fast multiplication \sep fast multi-point evaluation \sep fast minimal subspace polynomial \sep fast decoding
\end{keyword}

\end{frontmatter}

\section{Introduction}

\emph{Linearized polynomials} \citep{ore1933special} are polynomials of the form
\begin{align*}
a = \sum\limits_{k} a_k x^{q^k}, \quad a_k \in \Fqm \text{ (finite field)},
\end{align*}
which possess a ring structure with ordinary addition and polynomial composition.
They are an important class of polynomials which are of theoretical interest \citep{evans1992linearized,wu2013linearized} and have many applications in coding theory \citep{Gabidulin_TheoryOfCodes_1985,silva2008rank}, dynamical systems \citep{cohen2000dynamics} and cryptography \citep{gabidulin1991ideals}.
Especially in coding theory, designing fast algorithms for certain operations on these polynomials is crucial since it directly determines the complexity of decoding Gabidulin codes, an important class of rank-metric codes.

The operations that we consider in this paper are multiplication, division, $q$-transform, computing minimal subspace polynomials, multi-point evaluation, and interpolation of linearized polynomials of degree at most $s$ over $\Fqm$.

In this section, we omit log factors using the $\BigOTtext{\cdot}$ notation. These factors can be found in the respective theorems or references. By $\omega \leq 3$, we denote the matrix multiplication exponent.

\subsection{Related Work}
\label{subsec:previous_work}

For $s\leq m$ and $m$ admitting a low-complexity normal basis of $\Fqm$ over $\Fq$, \citet{SilvaKschischang-FastEncodingDecodingGabidulin-2009} and \citet{WachterAfanSido-FastDecGabidulin_DCC} presented algorithms for the $q$-transform with respect to a basis of $\Fqs \subseteq \Fqm$ ($\BigOtext{ms^2}$ over~$\Fq$), multi-point evaluation ($\BigOtext{m^2s}$ over~$\Fq$), and multiplication of linearized polynomials modulo $x^{q^m}-x$ ($\BigOtext{m^2s}$ over~$\Fq$), where the complexity bottleneck of the latter two methods is the so-called $q$-transform with respect to a basis of $\Fqm$ with complexity $\BigOtext{m^2s}$.

For arbitrary $s >0$, \citet[Sec.~3.1.2]{wachter2013decoding} presented an algorithm for multiplying two linearized polynomials of degree at most $s$ with complexity $\BigOtext{s^{\polymulexponent}}$ over $\Fqm$, where $\omega$ is the matrix multiplication exponent.
Finding a minimal subspace polynomial and performing a multi-point evaluation are both known to be implementable with $\BigOtext{s^2}$ operations in $\Fqm$, see~\citep{li2014transform}. Similarly, the known implementations of the $q$-transform require $\BigOtext{s^2}$ operations over $\Fqm$ \citep{wachter2013decoding}, and the interpolation $\BigOtext{s^3}$ over $\Fqm$ \citep{silva2007rank}.

Recently, \citet{caruso2017new} proposed algorithms for multiplication and division of skew polynomials (a generalization of linearized polynomials) that have complexity $\BigOTtext{s m}$. If $m \in o(s)$, then these algorithms are sub-quadratic in $s$. 
Further, they presented two Karatsuba-based algorithms where the so-called Karatsuba method has complexity $\BigOtext{s^{1.58} m^{1.41}}$ over $\Fq$, if $s>m$ and the so-called matrix method has complexity $\BigOtext{s^{1.58} m^{0.2}}$ over $\Fq$, if $s>m^2/2$.
For $m \in \Omega(s)$, it has been an open problem if division algorithms of sub-quadratic complexity exist.

Since operations over $\Fqm$ can be performed in $\BigOTtext{m}$ operations over $\Fq$ (cf.~\citep{couveignes2009elliptic}), a quadratic complexity $\BigOtext{s^2}$ over $\Fqm$ corresponds to $\BigOTtext{m s^2}$ over $\Fq$.
Hence, all the mentioned results over $\Fqm$ are not slower than the ones over $\Fq$ from \citep{SilvaKschischang-FastEncodingDecodingGabidulin-2009} and \citep{WachterAfanSido-FastDecGabidulin_DCC} and it suffices to compare our results to the cost bounds over $\Fqm$.

The results of this paper were partly presented at the IEEE International Symposium on Information Theory \citep{PuchingerWachterzeh-ISIT2016}, with an emphasis on the implications for coding theory and omitting many proofs and comparisons.

\subsection{Our Contribution}

In this paper, we present algorithms for the above operations that are sub-quadratic in the polynomial degree $s$ of the involved polynomials, independent of the field extension degree $m$.

First, we generalize the multiplication algorithm for linearized polynomials from \citep{wachter2013decoding}, which is based on a fragmentation of the involved polynomials similar to \citep{brent1980fast}, to the more general class of skew polynomials. We also analyze the resulting cost bounds in more details than in \citep{wachter2013decoding}.
This algorithm has complexity $\BigOtext{s^{\polymulexponent}}$ and, together with a result of \citet{caruso2017new}, implies a division algorithm in $\BigOTtext{s^{\polymulexponent}}$.

We show that computing the $q$-transform and its inverse can be reduced to a matrix-vector multiplication and solving a system of equations, respectively, where in both cases the involved matrix has Toeplitz form. Thus, it can be computed in $\BigOTtext{s}$ operations over $\Fqm$.

Our fast algorithms for multi-point evaluation and computing minimal subspace polynomials are divide-\&-conquer methods that call each other recursively. These convoluted calls enable us to circumvent problems that arise from the non-commutativity of the linearized polynomial multiplication.
We also propose a divide-\&-conquer interpolation algorithm that uses the new multi-point evaluation and minimal subspace polynomial routines.
All three methods use ideas from well-known algorithms from \citep[Section~10.1-10.2]{von2013modern} and can be implemented in $\BigOTtext{s^{\max\left\{\log_2(3), \polymulexponent\right\}}}$ operations over $\Fqm$.

Table~\ref{tab:overview_operations_linearized_new} summarizes the new cost bounds of operations that we prove in this paper.

Since fast multiplication directly determines the cost of the other algorithms in this paper, we present a lower bound on the complexity of multiplying two linearized polynomials of degree $s=m$ by showing that matrix multiplication can be reduced to the latter if elliptic normal bases of degree $m$ exist. The resulting bound implies that there cannot be a quasi-linear algorithm for multiplying linearized polynomials, independent of $m$, unless a quasi-quadratic matrix multiplication algorithm exists.

Finally, we use the results to derive a new cost bound, $\BigOTtext{n^{\max\left\{\log_2(3), \polymulexponent\right\}}}$, for decoding Gabidulin codes of length $n$. This is the first bound that is sub-quadratic in $n$.

\begin{table}[htb!]
	\caption{New and previous cost bounds over $\Fqm$ (see Section~\ref{subsec:previous_work} why known cost bounds over $\Fq$ are not tighter) for operations with linearized polynomialsof degree at most $s$ and coefficients in $\Fqm$. See \cref{subsec:linearized_polynomials} for a formal description of the operations.
	}
	\label{tab:overview_operations_linearized_new}
	\centering
	\begin{tabular}{p{0.55\textwidth} p{0.27\textwidth} p{0.06\textwidth}}
		\toprule	
		Operation (Source) & New	& Before \\
		\midrule
		Division (\citep{caruso2017new} and \cref{thm:division-compl})
		&
		$\BigOT{\min\left\{s m, s^{\polymulexponent} \right\}}$
		&
		$\BigOT{sm}$\\[1ex]
		(Inverse) $q$-Transform (\cref{thm:qt-compl})
		&
		$\BigOT{s}$
		&
		$\BigO{s^2}$\\[1ex]
		Minimal Subspace Polynomial Computation (\cref{thm:mspmpe-compl})
		&
		$\BigOT{s^{\max\left\{\log_2(3), \polymulexponent\right\}}}$
		&
		$\BigO{s^2}$ \\[1ex]
		Multi-point Evaluation (\cref{thm:mspmpe-compl})
		&
		$\BigOT{s^{\max\left\{\log_2(3), \polymulexponent\right\}}}$
		&
		$\BigO{s^2}$\\[1ex]
		Interpolation (\cref{thm:interpol-comp})
		&
		$\BigOT{s^{\max\left\{\log_2(3), \polymulexponent\right\}}}$
		&
		$\BigO{s^3}$\\
		\bottomrule
	\end{tabular}
\end{table}

\subsection{Structure of the Paper}

This paper is structured as follows.
In Section~\ref{sec:preliminaries}, we give definitions and formally introduce the operations that are considered in this paper.
Section~\ref{sec:fast_algos} contains the main results of the paper: we present fast algorithms for division, $q$-transform, calculation of minimal subspace polynomials, multi-point evaluation, and interpolation.
Using these new algorithms, we accelerate a known linearized polynomial multiplication algorithm and prove its optimality for the case $s=m$ in Section~\ref{sec:implications_and_optimality}.
In Section~\ref{sec:decoding}, we show how our fast algorithms for linearized polynomials imply sub-quadratic decoding algorithms for a special class of rank-metric codes, Gabidulin codes, and Section~\ref{sec:conclusion} concludes this paper.

\section{Preliminaries}\label{sec:preliminaries}

Let $q$ be a prime power, $\Fq$ be a finite field with $q$ elements and $\Fqm$ an extension field of $\Fq$. The field $\Fqm$ can be seen as an $m$-dimensional vector space over $\Fq$.
A subspace of $\Fqm$ is always meant w.r.t.~$\Fq$ as the base field.
For a given subset $A \subseteq \Fqm$, the subspace $\LH{A}$ is the $\Fq$-span of~$A$.

\subsection{Normal Bases}\label{subsec:normal_bases}

Normal bases facilitate calculations in finite fields and can therefore be used to reduce the computational complexity.
We shortly summarize important properties of normal bases in the following, cf.~\citep{Gao_NormalBases_1993,lidl1997finite,Menezes2010Applications}.
A basis $\Basis =\lbrace \beta_0, \beta_1,\dots, \beta_{m-1}\rbrace$ of $\Fqm$ over $\Fq$ is a \textit{normal basis} if $\beta_i= \Normelement^{[i]}$ for all $i$, where $\Normelement\in \mathbb{F}_{q^m}$ is called \emph{normal element}.
As shown in \citep[Thm.~2.35]{lidl1997finite}, 
there is a normal basis for any finite extension field $\Fqm$ over $\Fq$.

The \emph{dual basis} $\Dualbasis$ of a basis $\Basis$ is needed to switch between a polynomial and its $q$-transform (cf.~\cref{subsec:fast_qtrans}).
For a given basis $\Basis$ of $\Fqm$ over $\Fq$, there is a unique dual basis $\Dualbasis$.
The dual basis of a normal basis is also a normal basis, cf.~\citep[Thm.~1.1]{Menezes2010Applications}.

If we represent elements of $\Fqm$ in a normal basis over $\Fq$, applying the Frobenius automorphism $\cdot^q$ to an element can be accomplished in $\BigO{1}$ operations over $\Fqm$ as follows.
Let $[A_1,\dots,A_m]^T \in \Fq^{m \times 1}$ be the vector representation of $a \in \Fqm$ in a normal basis.
Then, for any~$j$, the vector representation of $a^{[j]}$ is given by $[A_{m-j}, A_{m-j+1}, \dots, A_0, A_1, \dots, A_{m-j-1}]^T$, which is just a cyclic shift of the representation of $a$.
The same holds for an arbitrary automorphism $\sigma \in \mathrm{Gal}(\Fqm/\Fq)$ since it is of the form $\sigma(\cdot) = \cdot^{q^i}$ for $i<m$.

\subsection{Linearized Polynomials}
\label{subsec:linearized_polynomials}

In this paper, we present operations with \emph{linearized polynomials}, also called $q$-polynomials and defined as follows.
A linearized polynomial \citep{ore1933special} is a polynomial of the form
\begin{equation*}
a = \sum\limits_{k=0}^{t} a_k x^{q^k} = \sum\limits_{k=0}^{t} a_k x^{[k]}, \quad a_k \in \Fqm, \quad t\in \mathbb{N},
\end{equation*}
where we use the notation $[i] := q^i$.
The set of all linearized polynomials for given $q$ and $m$ is denoted by $\Lset$.
We define the addition $+$ of $a,b \in \Lset$ as for ordinary polynomials
\begin{equation*}
a+b = \sum\limits_{i} (a_i+b_i) x^{[i]}
\end{equation*}
and the multiplication $\mul$ as
\begin{equation}\label{eq:lin-mult}
a \mul b = \sum\limits_{i} \left(\sum\limits_{j=0}^{i} a_j b_{i-j}^{[j]}\right) x^{[i]}.
\end{equation}
Note that if $\Lset$ is seen as a subset of $\Fqm[x]$, the multiplication $\mul$ equals the composition $a(b(x))$.
Using these operations, $(\Lset,+,\mul)$ is a (non-commutative) ring \citep{ore1933special}.
The identity element of $(\Lset,+,\mul)$ is $x^{[0]}=x$. In the following, all polynomials are linearized polynomials.

We say that $a \in \Lset$ has $q$-degree $\qdeg a = \max\{k \in \mathbb{N} : a_k \neq 0\}$, where $\max \emptyset := -\infty$. For $s \in \N$, we define the set $\Lsetmaxs := \{a \in \Lset \, : \, \qdeg a \leq s\}$, and $\Lsmallers$ analogously.
A polynomial $a$ is called monic if $a_{\qdeg a} = 1$.
Further, $\qdeg(a \mul b) = \qdeg a + \qdeg b$ and $\qdeg(a+b) \leq \max\{\qdeg a, \qdeg b\}$.

For $a \in \Lset$, the evaluation \citep[\emph{Operator Evaluation}]{boucher2014linear} is defined by
\begin{align*}
a(\cdot) \, : \, \Fqm \to \Fqm \; , \; \alpha \mapsto a(\alpha) = \sum\limits_{i} a_i \alpha^{[i]}.
\end{align*}
Since $\autom(\alpha) = \alpha^{q}$ is the Frobenius automorphism, $\alpha^{[i]} = \autom^{i}(\alpha)$ is also an automorphism and it can be shown that $a(\cdot)$ is an $\Fq$-linear map for any $a \in \Lset$.
It follows that the root space $\RSpace{a} = \{\alpha \in \Fqm \, : \, a(\alpha) = 0\}$ is a subspace of $\Fqm$.
It is also clear that $(a \mul b)(\alpha) = a(b(\alpha))$.

\subsubsection{Division}

The ring of linearized polynomials is a left and right Euclidean domain and therefore admits a left and right division.

\begin{lemma}[\citet{ore1933special}]\label{lem:division}
For all $a,b \in \Lset$, $b\neq 0$, there are unique $\qR,\qL \in \Lset$ (quotients) and $\rR,\rL \in \Lset$ (remainders) such that $\qdeg \rR  < \qdeg b$, $\qdeg \rL < \qdeg b$, and
\begin{align*}
a &= \qR \mul b + \rR \quad \text{(right division)}, \\
a &= b \mul \qL + \rL \quad \text{(left division)}.
\end{align*}
\end{lemma}

Lemma \ref{lem:division} allows us to define a (right) modulo operation on $\Lset$ such that $a \equiv b \mod c$ if there is a $d \in \Lset$ such that $a = b + d \mul c$.
In the following, we use this definition of ``mod''.

\subsubsection{Minimal Subspace Polynomials}
Subspace polynomials are special linearized polynomials, with the property that their $q$-degree equals their number of linearly independent roots.
\begin{lemma}[\citet{{lidl1997finite}}]\label{lem:MSP}
Let $\U$ be a subspace of $\Fqm$.
Then there exists a unique nonzero monic polynomial $\MSP{\U} \in \Lset$ of minimal degree such that $\ker(\MSP{\U}) = \U$. Its degree is $\qdeg \MSP{\U} =\dim \U$.
\end{lemma}

The polynomial  $\MSP{\U}$ in Lemma~\ref{lem:MSP} is called \emph{minimal subspace polynomial} (MSP) of $\U$.

\subsubsection{Multi-point Evaluation}

Multi-point evaluation (MPE) is the process of evaluating a polynomial $a \in \Lset$ at multiple points $\alpha_1,\dots,\alpha_s \in \Fqm$, i.e. computing the vector $\big[ a(\alpha_1), \dots, a(\alpha_s) \big]
\in \Fqm^s$.

Notice that for linearized polynomials $a(\beta_1\alpha_1 + \beta_2\alpha_2) = \beta_1 a(\alpha_1) + \beta_2a(\alpha_2)$ for any $\beta_1,\beta_2 \in \Fq$ and $\alpha_1, \alpha_2 \in \Fqm$. If we have therefore evaluated $a(x)$ at a few linearly independent points, the evaluation of any $\Fq$-linear combination of these points can be calculated by simple additions with almost no cost.

\subsubsection{Interpolation}
\label{subsec:MPE_IP_Intro}

The dual problem of MPE is to find a polynomial of bounded degree that evaluates at given distinct points to certain values and is called interpolation. It is based on the following lemma.

\begin{lemma}[{\citet[Sec. III-A]{silva2007rank}}]\label{lem:interpolation_existence}
Let $(x_1,y_1),\dots,(x_s,y_s) \in \Fqm^2$ such that $x_1,\dots,x_s$ are linearly independent over $\Fq$. Then there exists a unique interpolation polynomial $\IP{\{(x_i,y_i)\}_{i=1}^{s}} \in \Lsmallers$ such that
\begin{align*}
\IP{\{(x_i,y_i)\}_{i=1}^{s}}(x_i) = y_i, \quad \forall \, i=1,\dots,s.
\end{align*}
\end{lemma}

\subsubsection{The $q$-transform}
Let $s$ divide $m$ and let $\B_N = \{\beta^{[0]}, \dots, \beta^{[s-1]}\}$ be a normal basis of $\F_{q^s} \subseteq \Fqm$ over $\Fq$.

\begin{definition}
The $q$-transform (w.r.t. $s$ and $\B_N$) is a mapping $\qtr{\cdot} \, : \, \Lsmallers \to \Lsmallers \, , \, a \mapsto \qtr{a}$ with
\begin{align}
\qtr{a}_j = a(\beta^{[j]}) = \sum\limits_{i=0}^{s-1} a_i \beta^{[i+j]}, \quad \forall j=0,\dots,s-1. \label{eq:qt}
\end{align}
\end{definition}
Given a dual normal basis $\B_N^\perp = \{{\beta^\perp}^{[0]}, \dots, {\beta^\perp}^{[s-1]}\}$ of~$\B_N$, the inverse $q$-tranform can be computed by
$a_i = \qtr{a}({\beta^\perp}^{[i]}) = \sum_{j=0}^{s-1} \qtr{a}_j {\beta^\perp}^{[j+i]}$ for all $i=0,\dots,s-1$,
cf.~\citep{SilvaKschischang-FastEncodingDecodingGabidulin-2009}. Thus, the $q$-transform is bijective.

\subsection{Skew Polynomials}\label{subsec:skew}

Let $\K$ be a field.
The ring of \emph{skew polynomials} $\K[x;\autom,\delta]$ over $\K$ with automorphism $\autom \in \mathrm{Gal}(\Fqm/\Fq)$ and derivation $\delta$, satisfying $\delta(a+b) = \delta(a)+\delta(b)$ and $\delta(ab) = \delta(a)b + \autom(a)\delta(b)$ for all $a,b \in \K$, is defined as the set of polynomials
$\sum_i a_i x^i$, $a_i \in \K$, 
with the multiplication rule
\begin{align*}
xa = \autom(a) x + \delta(a) \quad \forall a \in \K
\end{align*}
and ordinary component-wise addition.
The degree of a skew polynomial is defined as usual.
$\K[x;\autom,\delta]$ is left and right Euclidean, i.e., Lemma \ref{lem:division} also holds for skew polynomials.
A comprehensive description of skew polynomial rings and their properties can be found in \citep{ore1933theory}.

In this paper, we only consider the special case $\delta = 0$, in which we abbreviate the ring by $\K[x; \autom]$.
Also, we restrict ourselves to finite fields $\K = \Fqm$.
Note that there is a ring isomorphism $\varphi : \Lset \to \Aset, \, \sum_{i} a_i x^{[i]} \mapsto \sum_{i} a_i x^i$, where $\autom(\cdot) = \cdot^q$ is the \emph{Frobenius automorphism}.
Although some of our results might extend to a broader class of skew polynomials, we consider $\Fqm[x;\autom]$ only as an auxiliary tool for obtaining fast algorithms for linearized polynomials.

\section{Fast Algorithms}\label{sec:fast_algos}

This section presents the main results of this paper: new fast algorithms for division (\cref{subsec:division}), $q$-transform (\cref{subsec:fast_qtrans}), calculation of the MSP for a given subspace (\cref{subsec:fast_msp}), multi-point evaluation (also in \cref{subsec:fast_msp}), and interpolation (\cref{subsec:fast_interp}).

\subsection{Notations}
We count complexities in terms of operations in the field $\Fqm$.
For convenience, we use the following notations.

\begin{definition}
Let $s \in \N$.
We define the (worst-case) complexity measures, i.e., the infimum of the worst-case complexities of algorithms that solve the given problem.
\begin{enumerate}[label=\roman*)]
\item Complexity of left- or right-dividing $a \in \Lsetmaxs$ by $b \in \Lsetmaxs$:
\begin{align*}
\ODiv{s}.
\end{align*}
\item Complexity of computing the MSP $\MSP{\LH{U}}$ for a generating set $U = \{u_1,\dots,u_s\}$ of a subspace of $\Fqm$:
\begin{align*}
\OMSP{s}.
\end{align*}
\item Complexity of a multi-point evaluation of $a \in \Lsetmaxs$ at the points $\alpha_1, \dots, \alpha_s \in \Fqm$:
\begin{align*}
\OMPE{s}.
\end{align*}
\item Complexity of computing the $q$-transform or its inverse of a polynomial $a \in \Lsmallers$, given a normal basis $\Normbasis = \{\Normelement^{[0]}, \dots, \Normelement^{[s-1]}\}$ of $\F_{q^s} \subseteq \Fqm$:
\begin{align*}
\OQT{s}.
\end{align*}
\item Complexity of finding the interpolation polynomial of $s$ point tuples
\begin{align*}
\OIP{s}.
\end{align*}
\end{enumerate}
\end{definition}

Table~\ref{tab:overview_operations_linearized_aw} summarizes best known cost bounds for these operations with linearized polynomials.
\begin{table}[htb!]
    \caption{Previously best known cost bounds over $\Fqm$ (see Section~\ref{subsec:previous_work} why known cost bounds over $\Fq$ are not tighter) for operations with linearized polynomials. For an overview of the results presented in this paper, see Table~\ref{tab:overview_operations_linearized_new}.}
    \label{tab:overview_operations_linearized_aw}
    \centering
	\begin{tabular}{p{0.14\linewidth} p{0.29\linewidth} p{0.47\linewidth}}
		\toprule	
		Operation	& Cost Bound	& Source \\
		\midrule
		$\ODiv{s}$
		&
		$\BigOT{s m}$
		&
		\citep{caruso2017new}\\[1ex]
		$\OMSP{s}$
		&
		$\BigO{s^2}$
		&
		\citep{silva2008rank} \\[1ex]
		$\OMPE{s}$
		&
		$\BigO{s^2}$
		&
		``naive'' ($s$ ordinary polynomial evaluations)\\[1ex]
		$\OQT{s}$
		&
		$\BigO{s^2}$
		&
		\citep{silva2008rank}\\[1ex]
		$\OIP{s}$
		&
		$\BigO{s^3}$
		&
		``naive'' (using linearized Lagrange bases, cf.~\citep{silva2007rank})\\
		\bottomrule
    \end{tabular}
\end{table}

\subsection{Fast Division}
\label{subsec:division}

In this section, we present a division algorithm that has sub-quadratic complexity in the polynomial degree $s$ for arbitrary $s$. The currently best-known division algorithm has complexity $\BigOTtext{s m}$ \citep{caruso2017new}, which is quasi-linear for $s \gg m$, but quadratic if $s \in \Theta(m)$. We improve upon this algorithm in the latter case.

Our method is based on the following result by \citet{caruso2017new}, which states that skew polynomial division can be reduced to multiplication in another skew polynomial ring.
We denote by $\OMulSkew{s}$ the complexity multiplying two skew polynomials in $\Fqm[x;\autom]_{\leq s}$.
\begin{theorem}[\citet{caruso2017new}]\label{thm:division-compl}
$\ODiv{s} \in \BigO{\OMulSkew{s} \log s}$.
\end{theorem}
We generalize the fast multiplication algorithm for linearized polynomials from \citep[Theorem~3.1]{wachter2013decoding} to arbitrary skew polynomial rings with derivation $\delta=0$ in order to obtain a sub-quadratic cost bound on $\OMulSkew{s}$.
The algorithm is based on a fragmentation of polynomials, which was used for calculating power series expansions in \citep{Brent_Kung_1978,paterson1973number} and is related to the baby-steps giant-steps method.

We say that polynomials $a,b \in \Aset$ overlap at $k$ positions if the intersection of their supports $\supp(a) := \{i : a_i \neq 0\}$ and $\supp(b) := \{i : b_i \neq 0\}$ has cardinality
\begin{align*}
|\supp(a) \cap \supp(b)| = k.
\end{align*}
If $k=0$, we say that $a$ and $b$ are non-overlapping. Obviously, the sum of two polynomials overlapping at $k$ (known) positions can be calculated with $k$ additions in $\Fqm$ when the overlapping positions are known.

\begin{theorem}\label{thm:multiplication_fragmentation}
Let $a,b \in \Asetmaxs$, $\sstar := \lceil \sqrt{s+1} \rceil$.
Then $c = a \amul b$ can be calculated in $\BigOtext{s^{\frac{3}{2}}}$ field operations, plus the cost of multiplying an $\sstar \times \sstar$ with an $\sstar \times (s+\sstar)$ matrix, using Algorithm~\ref{alg:Mul}.
\end{theorem}

\begin{pf}
We can fragment $a$ into $\sstar$ non-overlapping polynomials $a^{(i)}$ as
\begin{align*}
a &= \sum\limits_{i=0}^{\sstar-1} a^{(i)} = \sum\limits_{i=0}^{\sstar-1} \left(\sum\limits_{j=0}^{\sstar-1} a_{i \sstar + j} x^{i \sstar + j} \right)
\end{align*}
and the result $c$ of the multiplication $c = a \amul b$ can also be fragmented as
\begin{align*}
c = a \amul b = \left(\sum\limits_{i=0}^{\sstar-1} a^{(i)} \right) \amul b
= \sum\limits_{i=0}^{\sstar-1} (a^{(i)} \amul b ) =: \sum\limits_{i=0}^{\sstar-1} c^{(i)}
\end{align*}
with
\begin{align*}
c^{(i)} &= \left( \sum\limits_{j=0}^{\sstar-1} a_{i \sstar + j} x^{i \sstar + j} \right) \amul \left( \sum\limits_{k=0}^{s} b_k x^{k} \right)
= \sum\limits_{j=0}^{\sstar-1} \sum\limits_{k=0}^{s} \left( a_{i \sstar + j} x^{i \sstar + j} \mul b_k x^{k} \right) 
= \sum\limits_{j=0}^{\sstar-1} \sum\limits_{k=0}^{s} \left( a_{i \sstar + j} \automi{i \sstar + j}(b_k) x^{i \sstar + j + k} \right) \\
&= \sum\limits_{h=0}^{s+\sstar-1} \left( \sum\limits_{j=0}^{h} a_{i \sstar + j} \automi{i \sstar + j}(b_{h-j}) \right) x^{i \sstar + h} =: \sum\limits_{h=0}^{s+\sstar-1} c_h^{(i)} x^{i \sstar + h}.
\end{align*}
Thus the $c^{(i)}$'s pairwise overlap at not more than $s$ positions, which we know.
In order to obtain the polynomials $c^{(i)}$, we can use
\begin{align*}
&\automi{-i \sstar}(c_h^{(i)}) = \\
&\quad \automi{-i \sstar}\left[\sum\limits_{j=0}^{h} a_{i \sstar + j} \automi{i \sstar + j}(b_{h-j})\right]
\overset{\automi{-i \sstar} \text{ aut.}}{=} \sum\limits_{j=0}^{h} \automi{-i \sstar}(a_{i \sstar + j}) \automi{-i \sstar + i \sstar + j}(b_{h-j})
= \sum\limits_{j=0}^{h} \automi{-i \sstar}(a_{i \sstar + j}) \automi{j}(b_{h-j}).
\end{align*}
We can write this expression as a vector multiplication
\begin{align*}
\begin{bmatrix}
\automi{-i \sstar}(a_{i \sstar})
& \dots
& \automi{-i \sstar}(a_{i \sstar+\sstar-1})
\end{bmatrix}
\cdot
\begin{bmatrix}
\automi{0}(b_{h}) &
\automi{1}(b_{h-1}) &
\dots &
\automi{h}(b_{0}) &
0 &
\dots &
0
\end{bmatrix}^T,
\end{align*}
where the left vector does not depend on $h$ and the right side is independent of $i$. Thus, we can write the computation of $\automi{-i \sstar}(c_h^{(i)})$ as a matrix multiplication $C = A \cdot B$ with
\begin{align}
C &= \Big[C_{ij} \Big]_{i=0,\dots,\sstar-1}^{j=0,\dots,s+\sstar-1},
\;  C_{ij} = \automi{-i \sstar}(c_j^{(i)}),  \notag \\
A &= \Big[A_{ij} \Big]_{i=0,\dots,\sstar-1}^{j=0,\dots,\sstar-1},
\quad \; A_{ij} = \automi{-i \sstar}(a_{i \sstar+j}), \label{eq:mul_matrices}\\
B &= \Big[B_{ij} \Big]_{i=0,\dots,\sstar-1}^{j=0,\dots,s+\sstar-1},
\; B_{ij} = \begin{cases}
\automi{j}(b_{i-j}), &0 \leq i-j \leq s, \\
0, &\text{else}.
\end{cases} \notag
\end{align}

Setting up the matrices $A$ and $B$ costs $\sstar \cdot s + \sstar \cdot \sstar \approx s^{\frac{3}{2}}$ many computations of automorphisms to $\Fqm$ elements.
Computing the matrix $C$ from $A$ and $B$ requires a multiplication of an $\sstar \times \sstar$ with an $\sstar \times (s+\sstar)$ matrix.
Extracting $c_j^{(i)} = \automi{i \sstar}(C_{ij})$ from $C$ costs a computation of an automorphism each, thus $\approx s^{\frac{3}{2}}$ computations in total.
In order to obtain the skew polynomial $c$, we need to add up the $c^{(i)}$'s. For some $k<\sstar$, the polynomials $\sum_{i=0}^{k-1} c^{(i)}$ and $c^{(k)}$ overlap at not more than $s$ positions.
Since we know these overlapping positions, we can compute the sum of all $c^{(i)}$'s in $\BigO{\sstar \cdot s} = \BigOtext{s^{\frac{3}{2}}}$ time.
In a finite field, the computation of an automorphism can be done in $\BigOtext{1}$, so Algorithm~\ref{alg:Mul} costs $\BigOtext{s^{\frac{3}{2}}}$, plus the matrix multiplication.
\end{pf}

\printalgoIEEE{
\DontPrintSemicolon
\KwIn{$a,b \in \Asetmaxs$}
\KwOut{$c = a \amul b$}
Set up matrices $A$ and $B$ as in~\eqref{eq:mul_matrices} \hfill \tcp{$s^{\frac{3}{2}} \cdot\BigO{1}$} \label{line:Mul_AB}
$C \gets A \cdot B$ \hfill \tcp{$\sstar \cdot \BigO{({\sstar})^\omega}$ or $(\sstar)^{3.2699}$}  \label{line:Mul_MatrixMul}
Extract the $c^{(i)}$'s from $C$ as in~\eqref{eq:mul_matrices} \hfill \tcp{$s^{\frac{3}{2}} \cdot\BigO{1}$}  \label{line:Mul_C}
\Return{$c \gets \sum_{i=0}^{\sstar-1} c^{(i)}$} \hfill \tcp{$\BigO{s^{\frac{3}{2}}}$}  \label{line:Mul_Add}
\caption{Multiplication}
\label{alg:Mul}
}

\begin{corollary}
\label{cor:multiplication}
Different techniques for the multiplication of the $\sstar \times \sstar$ with the $\sstar \times (s+\sstar)$ matrices in Theorem~\ref{thm:multiplication_fragmentation} result in the following cost bounds on the multiplication of skew polynomials:
\begin{enumerate}[label=\roman*)]
\item Using $\sstar+1$ many multiplications of $\sstar \times \sstar$ with $\sstar \times \sstar$ matrices, we obtain
\begin{align*}
\OMulSkew{s} \in \BigO{\sstar \cdot (\sstar)^\omega} \subseteq \BigO{s^{\frac{\omega+1}{2}}}.
\end{align*}
For instance, we get
\begin{align*}
\OMulSkew{s} \in
\begin{cases}
\BigO{s^{1.91}}, &\omega \approx 2.8074 \text{ \citep{strassen1969gaussian}},\\
\BigO{s^{1.69}}, &\omega \approx 2.376 \text{ \citep{coppersmith1990matrix}}.
\end{cases}
\end{align*}
\item Direct multiplication algorithms for rectangular matrices (cf.~\citep{huang1998fast,ke2008fast}) result in
\begin{align*}
\OMulSkew{s} \in \BigO{(\sstar)^{3.2699}} \subseteq \BigO{s^{1.635}},
\end{align*}
where the power $3.2699$ \citep[Example~1]{ke2008fast} holds for multiplying an $\sstar \times \sstar$ with an $\approx \sstar \times (\sstar)^2$ matrix.
\end{enumerate}
\end{corollary}

\begin{remark}\label{rem:pracical_complexities}
Naive skew/linearized polynomial multiplication using the definition from~\eqref{eq:lin-mult} uses approximately $2s^2$ many field operations. For comparison, if we use case $\mathrm{(i)}$ of Corollary~\ref{cor:multiplication} with naive matrix multiplication, where each multiplication uses approximately $2(\sstar)^3$ operations, skew polynomial multiplication takes $\approx \sstar \cdot (2(\sstar)^3) = 2 (\sstar)^4 = 2 s^2$ operations in total.
Thus, we improve upon the naive case as soon as the algorithm for multiplying two matrices of dimension $\sstar \times \sstar$ is faster than $2(\sstar)^3$.

For instance, the algorithm of \citet{strassen1969gaussian} uses $\approx 4.7 (\sstar)^{\log_2(7)}$ field operations, which is smaller than $2(\sstar)^3$ for $\sstar \geq 85$, or in other words $s \geq 7225$. The algorithm by \citet{coppersmith1990matrix} has a much larger ``hidden constant'' and improves upon the naive case only for huge values of $s$.

Besides the asymptotic improvement, Algorithm~\ref{alg:Mul} can yield a practical speed-up, compared to a naive implementation that does not use linear-algebraic operations, by relying on efficiently implemented linear algebra libraries that are optimized for the used programming language or hardware.
\end{remark}

Using Theorem~\ref{thm:division-compl} and Corollary~\ref{cor:multiplication}, we obtain the following new cost bound on the division of linearized polynomials.
\begin{corollary}\label{cor:division-compl-linearized}
$\ODiv{s} \in \BigO{s^{\polymulexponent} \log s}$.
\end{corollary}

As a direct consequence of the result above, we obtain a fast (half) linearized extended Euclidean algorithm (LEEA) (cf.~\citep[Corollary~3.2]{wachter2013decoding}).
\begin{corollary}\label{cor:compl-leea}
The fast (half) LEEA from \citep[Algorithm~3.4]{wachter2013decoding} for two input polynomials $a,b$, where $s:=\deg_q a \geq \deg_q b$ can be implemented in
\begin{align*}
\BigO{\max\left\{\ODiv{s},\OMulSkew{s}\right\}\log s} \subseteq \BigO{s^{\polymulexponent}\log^2(s)}
\end{align*}
operations over $\Fqm$.
\end{corollary}

\subsection{Fast $q$-Transform}\label{subsec:fast_qtrans}

The following theorem states that both the $q$-transform and its inverse can be obtained in quasi-linear time over $\Fqm$.
Recall that $s$ must divide $m$ in order for the $q$-transform to be well-defined.
The idea of the fast $q$-transform is based on the fact that the $q$-transform is basically the multiplication of the vector with a Toeplitz matrix. Since Toeplitz matrix multiplication can be reduced to multiplication of polynomials in $\Fqm[x]$ (cf.~\citep{von2013modern}), also the $q$-transform can be implemented in quasi-linear time over $\Fqm$.

\begin{theorem}\label{thm:qt-compl}
$\OQT{s} \in \BigO{s \log^2(s) \log (\log(s))}$.
\end{theorem}

\begin{pf}
Let $a \in \Lsmallers$. From \eqref{eq:qt} we know that
\begin{align*}
(\qtr{a}_0, \dots, \qtr{a}_{s-1})
&= (a_{s-1}, \dots, a_0) \cdot
\underset{=: B}{
\underbrace{
\begin{bmatrix}
\beta^{[s-1]} & \beta^{[s]} 	& \beta^{[s+1]} & \dots  & \beta^{[2s-1]} \\
\beta^{[s-2]} & \beta^{[s-1]}	& \beta^{[s]} 	& \dots  & \beta^{[2s-2]} \\
\vdots        & \vdots      & \vdots      	& \ddots & \vdots       	  \\
\beta^{[0]}   & \beta^{[1]} & \beta^{[2]} 	& \dots  & \beta^{[s-1]}
\end{bmatrix},
}}
\end{align*}
where the matrix $B$ is an $s \times s$ Toeplitz matrix over $\Fqm$.
At the same time, it is a $q$-Vandermonde matrix which is invertible (see~\citep[Lemma~3.5.1]{lidl1997finite}).

As described in \citep[Problems~2.5.1]{bini2012polynomial}, Toeplitz matrix vector multiplication can be reduced to multiplication of $\Fqm[x]$ polynomials with degree $\leq s$, which has complexity $\BigO{s \log(s) \log (\log(s))}$, cf.~\citep{von2013modern}.

The inverse $q$-transform consists of solving a Toeplitz linear system, which is reducible to a Pad\'e approximation problem, that again can be solved using the extended Euclidean algorithm over $\Fqm[x]$, cf.~\citep{brent1980fast}.
A fast Extended Euclidean Algorithm with stopping condition was introduced by Aho and Hopcroft in \citep{aho_hopcroft_fast_algorithms}. Its complexity is shown to be $\mathcal{O}(\mathcal{\tilde{M}}(s) \log(s))$, where $\tilde{\mathcal{M}}(s)$ is the complexity of multiplying two polynomials of degree $s$ in $\Fqm[x]$. However, for some special cases the algorithm does not work properly and therefore the improvements from \citep{GustavsonYun1979}, \citep{BrentGustavsonYun1980} have to be considered. This fast EEA was summarized and proven in \citep{blahut_fast_algorithms}.
The resulting complexity of solving the Toeplitz linear system is thus $\BigO{s \log^2(s) \log (\log(s))}$.
\end{pf}

\subsection{Fast Computation of MSP and MPE}\label{subsec:fast_msp}

In this subsection, we consider an efficient way to calculate the minimal subspace polynomial and the multi-point evaluation.
Fast algorithms for multi-point evaluation at a set $S \subseteq \Fqm$ over $\Fqm[x]$ typically pre-compute a sub-product tree (consisting of polynomials of the form $M_U = \prod_{u \in U} (x-u)$ for $U \subseteq S$) and then use divide-and-conquer methods for fast MPE.
Such a sub-product tree can only be computed fast since in the commutative case, the polynomial $M_U$ can be written as the product of two such polynomials of a partition $U = A \cup B$, $A \cap B = \emptyset$,
\begin{align*}
M_U = M_A \cdot M_B.
\end{align*}
The equivalent statement for linearized polynomials, using minimal subspace polynomials, is given in Lemma~\ref{lem:MSP_recursion} (see below).
In contrast to the commutative case, one of the factors depends on a multi-point evaluation of the other factor.
Hence, we cannot immediately apply the known methods.

The following two lemmas lay the foundation to algorithms that compute MSPs and MPEs by convoluted recursive calls of each other. Thus, we need to analyze their complexities jointly.

\begin{lemma}[\citet{li2014transform}]\label{lem:MSP_recursion}
Let $U = \{u_1,\dots,u_s\}$ be a generating set of a subspace $\U \subseteq \Fqm$, $A,B \subseteq \Fqm$ such that $U = A \cup B$. Then,
\begin{align*}
\MSP{\U} = \MSP{\LH{U}} = \MSP{\LH{\MSP{\LH{A}}(B)}} \mul \MSP{\LH{A}}
\end{align*}
and
\begin{align}
M_{\LH{u_i}} = \begin{cases}
x^{[0]}, & \text{if } u_i=0, \\
x^{[1]} - u_i^{q-1} x^{[0]}, & \text{else.}
\end{cases} \label{eq:MSP_recursion_base}
\end{align}
\end{lemma}

\begin{lemma}\label{lem:MPE_recursion}
Let $a \in \Lset$ and let $U, A,B \subseteq \Fqm$ where $A,B\subseteq \Fqm$ are disjoint and $U = A \cup B$. 
Let $\remainder_A,\remainder_B$ be the remainders of the right divisions of $a$ by $\MSP{\LH{A}}$ and $\MSP{\LH{B}}$ respectively.\\
Then, the multi-point evaluation of $a$ at the set $U$ is
\begin{align*}
a(U) = \remainder_A(A) \cup \remainder_B(B).
\end{align*}
If $U = \{u\}$ and $\qdeg a \leq 1$,
\begin{align}
a(U) = \{a(u) = a_1 u^{[1]} + a_0 u^{[0]} \}. \label{eq:MPE_recursion_base}
\end{align}
\end{lemma}

\begin{pf}
Let $u \in U$. If $u \in A$,
\begin{align*}
a(u) &= (\quo_A \mul \MSP{\LH{A}} + \remainder_A)(u) = \quo_A(\underset{=0}{\underbrace{\MSP{\LH{A}}(u)}}) + \remainder_A(u)
= \underset{=0}{\underbrace{\quo_A(0)}} + \remainder_A(u) = \remainder_A(u).
\end{align*}
Otherwise, $u \in B$ and
\begin{align*}
a(u) &= (\quo_B \mul \MSP{\LH{B}} + \remainder_B)(u) = \quo_B(\underset{=0}{\underbrace{\MSP{\LH{B}}(u)}}) + \remainder_B(u)
= \underset{=0}{\underbrace{\quo_B(0)}} + \remainder_B(u) = \remainder_B(u).
\end{align*}
Thus, $a(U) = \remainder_A(A) \cup \remainder_B(B)$.
Equation \eqref{eq:MPE_recursion_base} follows directly from the definition of the evaluation map.
\end{pf}

This yields the main statement of this subsection.
\begin{theorem}\label{thm:mspmpe-compl}
Finding the MSP of an $\Fqm$-subspace spanned by $s$ elements of $\Fqm$ and computing the MPE of a polynomial of $q$-degree at most $s$ at $s$ many points can be implemented in
\begin{align*}
\OMSP{s}, \OMPE{s} &\in \BigO{\max\left\{s^{\log_2(3)} \log(s), \OMulSkew{s}, \ODiv{s}\right\}} \\
&\subseteq \BigO{s^{\max\left\{\log_2(3), \polymulexponent \right\}} \log(s)}.
\end{align*}
operations over $\Fqm$ using Algorithms~\ref{alg:MSP} and \ref{alg:MPE} respectively.
\end{theorem}

\begin{pf}
We prove that Algorithm \ref{alg:MSP} for computing the MSP and Algorithm \ref{alg:MPE} for MPE are correct and have the desired complexity.

{\bf Correctness:}
Since the algorithms call each other recursively, we need to prove their correctness jointly by induction.

For $s=1$, Algorithm~\ref{alg:MSP} returns the base case of Lemma~\ref{lem:MSP_recursion} and Algorithm~\ref{alg:MPE} uses Equation~\eqref{eq:MPE_recursion_base} of Lemma~\ref{lem:MPE_recursion} to compute the evaluation of a polynomial of $\qdeg a \leq 1$ at one point.

Now suppose that both algorithms work up to an input size of $s-1$ for some $s \geq 2$.
Then, Algorithm~\ref{alg:MSP} works for an input of size $s$ because it uses the recursion formula of Lemma~\ref{lem:MSP_recursion} to reduce the problem to two MSP computations and a multi-point evaluation, each of input size $\approx s/2 \leq s-1$.
Algorithm~\ref{alg:MPE} works for an input of size $s$ due to Lemma~\ref{lem:MPE_recursion} and a similar argument.

Hence, both algorithms are correct.

{\bf Complexity:}

The lines of Algorithm~\ref{alg:MSP} have the following complexities:
\begin{itemize}
\item The complexities of Lines \ref{line:MSP_base_case} (base case) and \ref{line:MSP_partition} (partitioning of $U$) are negliglible.
\item Lines \ref{line:MSP_MSPA} and \ref{line:MSP_MSPB} both have complexity $\OMSP{\frac{s}{2}}$ because $|A| \approx |B| \approx \frac{|U|}{2} = \frac{s}{2}$.
\item Line \ref{line:MSP_MPEB} computes the result in $\OMPE{\frac{s}{2}}$ time because $\qdeg \MSP{\langle A \rangle} \lesssim |B| \approx \frac{|U|}{2} = \frac{s}{2}$.
\item Line \ref{line:MSP_mul} performs a multiplication of two polynomials of $q$-degree $\lesssim \frac{s}{2} \leq s$ and has time complexity $\OMul{s}$.
\end{itemize}
In total, we obtain
\begin{align}
\OMSP{s} = 2 \cdot \OMSP{\tfrac{s}{2}} + \OMPE{\tfrac{s}{2}} + \OMulSkew{s}. \label{eq:OMSP_recursion}
\end{align}

Algorithm~\ref{alg:MPE} consists of the following steps:
\begin{itemize}
\item Again, the complexities of Lines \ref{line:MPE_base_case} (base case) and \ref{line:MPE_partition} (partitioning of $U$) are negliglible.
\item Lines \ref{line:MPE_MSPA} and \ref{line:MPE_MSPB} compute the MSP of bases with input size $|A| \approx |B| \approx \frac{|U|}{2} = \frac{s}{2}$, so both have complexity $\OMSP{\frac{s}{2}}$.
\item Lines \ref{line:MPE_div_A} and \ref{line:MPE_div_B} divide polynomials from $\Lsetmaxs$ and therefore have complexity $\ODiv{s}$ each.
\item Line \ref{line:MPE_MPE} performs two multi-point evaluations of polynomials with $q$-degree $< |A| \approx |B| \approx \frac{s}{2}$ (cf. Lemma \ref{lem:division}, $\qdeg$ of remainder) at $|A| \approx |B| \approx \frac{s}{2}$ positions. Thus, the line has complexity $2 \cdot \OMPE{\frac{s}{2}}$.
\end{itemize}
Summarized, we get
\begin{align}
\OMPE{s} = 2 \cdot \OMSP{\tfrac{s}{2}} + 2 \cdot \OMPE{\tfrac{s}{2}} + 2 \cdot \ODiv{s}. \label{eq:OMPE_recursion_bad}
\end{align}
In fact, the MSP computed in Line \ref{line:MPE_MSPA} of Algorithm \ref{alg:MPE} is the same as the MSP which was computed in Line \ref{line:MSP_MSPA} of Algorithm~\ref{alg:MSP} at the same recursion depth before (note that $\AMSP{U}$ first calls $\AMSP{A}$ and then $\AMPE{\MSP{\LH{A}}}{B}$).
This means that we can store this polynomial instead of recomputing it and can reduce~\eqref{eq:OMPE_recursion_bad} to 
\begin{align}
\OMPE{s} = \OMSP{\tfrac{s}{2}} + 2 \cdot \OMPE{\tfrac{s}{2}} + 2 \cdot \ODiv{s}. \label{eq:OMPE_recursion}
\end{align}
We define $C(s) := \max\left\{\OMPE{s}, \OMSP{s}\right\}$ and derive an upper bound on $C(s)$. Using \eqref{eq:OMSP_recursion} and \eqref{eq:OMPE_recursion}, we obtain
\begin{align*}
C(s) \leq 3 \cdot C\left(\tfrac{s}{2}\right) + \max\left\{\OMulSkew{s},  2 \cdot \ODiv{s}\right\}. 
\end{align*}
We distinguish three cases and use the master theorem:
\begin{itemize}
\item If $\max\left\{\OMulSkew{s}, \ODiv{s}\right\} \in \BigO{s^{\log_2(3)-\varepsilon}}$ for some $\varepsilon>0$, then
\begin{align*}
C(s) \in \BigO{s^{\log_2(3)}}.
\end{align*}
\item If $\max\left\{\OMulSkew{s}, \ODiv{s}\right\} \in \BigTheta{s^{\log_2(3)}}$, then
\begin{align*}
C(s) \in \BigO{s^{\log_2(3)} \log(s)}.
\end{align*}
\item If $\max\left\{\OMulSkew{s}, \ODiv{s}\right\} \in \BigOmega{s^{\log_2(3)+\varepsilon}}$ for some $\varepsilon>0$, then
\begin{align*}
C(s) \in \BigO{\max\left\{\OMulSkew{s}, \ODiv{s}\right\}}.
\end{align*}
\end{itemize}
In summary, we obtain
\begin{align*}
\OMSP{s}, \OMPE{s} &\in \BigO{\max\left\{s^{\log_2(3)} \log(s), \OMulSkew{s}, \ODiv{s}\right\}} \\
&\subseteq \BigO{s^{\max\left\{\log_2(3), \polymulexponent \right\}} \log(s)}.
\end{align*}
\end{pf}

\printalgoIEEE{
\DontPrintSemicolon
\KwIn{Generating set $U = \{u_1, \dots, u_s\}$ of a subspace $\U \subseteq \Fqm$.}
\KwOut{MSP $\MSP{\langle U \rangle}$.}
\If{$s=1$}{\Return{$M_{\langle u_1\rangle}(x)$ according to \eqref{eq:MSP_recursion_base}} \hfill \tcp{$\BigO{1}$}} \label{line:MSP_base_case}
\Else{
$A \gets \{u_1,\dots,u_{\gauss{\frac{s}{2}}}\}$, $B \gets \{u_{\gauss{\frac{s}{2}}+1},\dots,u_s\}$ \hfill \tcp{$\BigO{1}$} \label{line:MSP_partition}
$\MSP{\langle A \rangle} \gets \AMSP{A}$ \hfill \tcp{$\OMSP{\tfrac{s}{2}}$}  \label{line:MSP_MSPA}
$\MSP{\langle A \rangle}(B) \gets \AMPE{\MSP{\langle A \rangle}}{B}$  \hfill \tcp{$\OMPE{\tfrac{s}{2}}$}  \label{line:MSP_MPEB}
$\MSP{\MSP{\langle A \rangle}(B) \rangle} \gets \AMSP{\MSP{\langle A \rangle}(B)}$ \hfill \tcp{$\OMSP{\tfrac{s}{2}}$}  \label{line:MSP_MSPB}
\Return{$\MSP{\MSP{\langle A \rangle}(B) \rangle} \mul \MSP{\langle A \rangle}$} \hfill \tcp{$\OMulSkew{s}$}  \label{line:MSP_mul}
}
\caption{$\AMSP{U}$}
\label{alg:MSP}
}

\printalgoIEEE{
\DontPrintSemicolon
\KwIn{$a\in \Lsetmaxs$, $\{u_1, \dots,u_s\} \in \Fqm^s$}
\KwOut{Evaluation of $a$ at all points $u_i$}
\If{$s=1$}{\Return{$\{a_1 u_1^{[1]} + a_0 u_1^{[0]}\}$} \hfill \tcp{$\BigO{1}$}} \label{line:MPE_base_case}
\Else{
$A\! \gets\!\! \{u_1,\dots,u_{\gauss{\frac{s}{2}}}\}$, $B \!\gets\!\! \{u_{\gauss{\frac{s}{2}}+1},\dots,u_s\}$ \hfill \tcp{$\!\!\!\BigO{1}$}\label{line:MPE_partition}
$\MSP{\langle A \rangle} \gets \AMSP{A}$ \hfill \tcp{$\OMSP{\tfrac{s}{2}}$} \label{line:MPE_MSPA}
$\MSP{\langle B \rangle} \gets \AMSP{B}$ \hfill \tcp{$\OMSP{\tfrac{s}{2}}$} \label{line:MPE_MSPB}
$[\quo_A, \remainder_A] \gets \ARDIV{a, \MSP{\langle A \rangle}}$ \hfill \tcp{$\ODiv{s}$} \label{line:MPE_div_A}
$[\quo_B, \remainder_B] \gets \ARDIV{a, \MSP{\langle B \rangle}}$ \hfill \tcp{$\ODiv{s}$} \label{line:MPE_div_B}
\Return{$\AMPE{\remainder_A}{A} \cup \AMPE{\remainder_B}{B}$} \hfill \tcp{$2 \cdot \OMPE{\tfrac{s}{2}}$} \label{line:MPE_MPE}
}
\caption{$\AMPE{a}{\{u_1, \dots,u_s\}}$}
\label{alg:MPE}
}

\subsection{Fast Interpolation}\label{subsec:fast_interp}

In this section, we present a fast divide-\&-conquer interpolation algorithm for linearized polynomials that relies on fast algorithms for both computing MSPs and MPEs. The idea resembles the fast interpolation algorithm in $\Fqm[x]$ from \cite[Section~10.2]{von2013modern} with additional considerations for the non-commutativity, and is based on the following lemma.

\begin{lemma}\label{lem:interpolation_recursion}
Let $(x_i,y_i)$ be as in Lemma~\ref{lem:interpolation_existence}. The interpolation polynomial fulfills
\begin{align*}
\IP{\{(x_i,y_i)\}_{i=1}^{s}} = \IP{\{(\tilde{x}_i,y_i)\}_{i=1}^{\gauss{\frac{s}{2}}}} \mul \MSP{\LH{x_{\gauss{\frac{s}{2}}+1}, \dots, x_s}}
+ \IP{\{(\tilde{x}_i,y_i)\}_{i=\gauss{\frac{s}{2}}+1}^{s}} \mul \MSP{\LH{x_1,\dots,x_{\gauss{\frac{s}{2}}}}}
\end{align*}
with
\vspace{-0.3cm}
\begin{align*}
\tilde{x}_i :=
\begin{cases}
\MSP{\LH{x_{\gauss{\frac{s}{2}}+1}, \dots, x_s}}(x_i), &\text{if } i=1,\dots,\gauss{\frac{s}{2}} \\
\MSP{\LH{x_1,\dots,x_{\gauss{\frac{s}{2}}}}}(x_i), &\text{otherwise}
\end{cases}
\end{align*}
and $\IP{\{(x_i,y_i)\}_{i=1}^{1}} = \frac{y_1}{x_1} x^{[0]}$ (base case $s=1$).
\end{lemma}

\begin{pf}
For $i=1,\dots,\gauss{\frac{s}{2}}$, the $\tilde{x}_i$ are linearly independent since the $x_i$ are linearly independent and $\MSP{\LH{x_{\gauss{\frac{s}{2}}+1}, \dots, x_s}}(\cdot)$ is a linear map whose kernel is spanned by $x_{\gauss{\frac{s}{2}}+1}, \dots, x_s$, and therefore does not include any $x_i$ for $i=1,\dots,\gauss{\frac{s}{2}}$.
Furthermore,
\vspace{-0.3cm}
\begin{align*}
\IP{\{(x_i,y_i)\}_{i=1}^{s}}(x_i) &= \IP{\{(\tilde{x}_i,y_i)\}_{i=1}^{\gauss{\frac{s}{2}}}} \big( \overset{= \tilde{x_i}}{\overbrace{\MSP{\LH{x_{\gauss{\frac{s}{2}}+1}, \dots, x_s}}(x_i)}} \big)
+ \IP{\{(\tilde{x}_i,y_i)\}_{i=\gauss{\frac{s}{2}}+1}^{s}} \big( \underset{= 0}{\underbrace{\MSP{\LH{x_1,\dots,x_{\gauss{\frac{s}{2}}}}}(x_i)}} \big) \\
&= \IP{\{(\tilde{x}_i,y_i)\}_{i=1}^{\gauss{\frac{s}{2}}}} (\tilde{x}_i) + \IP{\{(\tilde{x}_i,y_i)\}_{i=\gauss{\frac{s}{2}}+1}^{s}}(0)
= y_i + 0 = y_i.
\end{align*}
By the same argument, also $\tilde{x}_{\gauss{\frac{s}{2}}+1}, \dots, \tilde{x}_s$ are linearly independent and
$\IP{\{(x_i,y_i)\}_{i=1}^{s}}(x_i) = y_i$ 
for all $i=\gauss{\frac{s}{2}}+1,\dots,s$.
Since in addition,
\begin{align*}
&\qdeg{\IP{\{(x_i,y_i)\}_{i=1}^{s}}} \leq \\
&\max\big\{ \underset{<\frac{s}{2}}{\underbrace{\qdeg \big(\IP{\{(\tilde{x}_i,y_i)\}_{i=1}^{\gauss{\frac{s}{2}}}}\big)}} + \underset{\leq\frac{s}{2}}{\underbrace{\qdeg \big( \MSP{\LH{x_{\gauss{\frac{s}{2}}+1}, \dots, x_s}} \big)}}, \underset{<\frac{s}{2}}{\underbrace{ \qdeg \big(\IP{\{(\tilde{x}_i,y_i)\}_{i=\gauss{\frac{s}{2}}+1}^{s}} \big)}} + \underset{\leq\frac{s}{2}}{\underbrace{\qdeg \big(\MSP{\LH{x_1,\dots,x_{\gauss{\frac{s}{2}}}}} \big) }} \big\}< s,
\end{align*}
the polynomial
$
\IP{\{(\tilde{x}_i,y_i)\}_{i=1}^{\gauss{\frac{s}{2}}}} \mul \MSP{\LH{x_{\gauss{\frac{s}{2}}+1}, \dots, x_s}}
+ \IP{\{(\tilde{x}_i,y_i)\}_{i=\gauss{\frac{s}{2}}+1}^{s}} \mul \MSP{\LH{x_1,\dots,x_{\gauss{\frac{s}{2}}}}}
$
is the desired interpolation polynomial $\IP{\{(x_i,y_i)\}_{i=1}^{s}}(x_i)$ of Lemma \ref{lem:interpolation_existence}.
The case $s=1$ is obvious.
\end{pf}

Lemma~\ref{lem:interpolation_recursion} implies a divide-\&-conquer interpolation strategy. The method is outlined in Algorithm~\ref{alg:IP}, whose complexity is stated in the following theorem.

\begin{theorem}\label{thm:interpol-comp}
Computing the interpolation polynomial of $s$ point tuples can be implemented in
\begin{align*}
\OIP{s} \in \BigOtext{\OMSP{s}} \subseteq \BigOtext{s^{\max\left\{\log_2(3), \polymulexponent \right\}} \log(s)}
\end{align*}
operations over $\Fqm$ using Algorithm~\ref{alg:IP}.
\end{theorem}

\begin{pf}
Algorithm~\ref{alg:IP} computes the correct interpolation polynomial due to Lemma~\ref{lem:interpolation_recursion}.
Its lines have the following complexities:
\begin{itemize}
\item Lines~\ref{line:IP_base_case} and \ref{line:IP_partition} are again negligible.
\item The complexities of Lines~\ref{line:IP_MSPA} and \ref{line:IP_MSPB} are $\OMSP{\tfrac{s}{2}}$.
\item Lines~\ref{line:IP_x_tilde_1} and \ref{line:IP_x_tilde_1} take $\OMPE{\tfrac{s}{2}}$ time each.
\item The algorithm calls itself recursively with input size $\approx \frac{s}{2}$ in Lines~\ref{line:IP_IP1} and \ref{line:IP_IP2}.
\item Finally, the result is reassembled in line~\ref{line:IP_mul} using two multiplications in $2 \cdot \OMulSkew{\tfrac{s}{2}}$ time.
\end{itemize}
Overall, we have
\begin{align*}
\OIP{s} &= 2 \cdot \OIP{\tfrac{s}{2}} + 2 \cdot \left(\OMSP{\tfrac{s}{2}} + \OMPE{\tfrac{s}{2}} + \OMulSkew{\tfrac{s}{2}}\right)
= 2 \cdot \OIP{\tfrac{s}{2}} + \BigO{\OMSP{s}}
\end{align*}
By the master theorem, we obtain the desired complexity $\OIP{s} \in \BigOtext{\OMSP{s}}$.
\end{pf}

\printalgoIEEE{
\DontPrintSemicolon
\KwIn{$(x_1,y_1),\dots,(x_s,y_s) \in \Fqm^2$, $x_i$ linearly independent}
\KwOut{Interpolation polynomial $\IP{\{(x_i,y_i)\}_{i=1}^{s}}$}
\If{$s=1$}{\Return{$\{\frac{y_1}{x_1} x^{[0]}\}$} \hfill \tcp{$\BigO{1}$}} \label{line:IP_base_case}
\Else{
$A \gets \{x_1,\dots,x_{\gauss{\frac{s}{2}}}\}$, $B \gets \{x_{\gauss{\frac{s}{2}}+1},\dots,x_s\}$ \hfill \tcp{$\BigO{1}$}\label{line:IP_partition}
$\MSP{\langle A \rangle} \gets \AMSP{A}$ \hfill \tcp{$\OMSP{\tfrac{s}{2}}$} \label{line:IP_MSPA}
$\MSP{\langle B \rangle} \gets \AMSP{B}$ \hfill \tcp{$\OMSP{\tfrac{s}{2}}$} \label{line:IP_MSPB}
$\{\tilde{x}_1,\dots,\tilde{x}_{\gauss{\frac{s}{2}}}\} \gets \AMPE{\MSP{\langle B \rangle}}{A}$ \hfill \tcp{$\OMPE{\tfrac{s}{2}}$} \label{line:IP_x_tilde_1}
$\{\tilde{x}_{\gauss{\frac{s}{2}}+1},\dots,\tilde{x}_s\} \gets \AMPE{\MSP{\langle A \rangle}}{B}$ \hfill \tcp{$\OMPE{\tfrac{s}{2}}$} \label{line:IP_x_tilde_2}
$\IP{1} \gets \AIP{\{(\tilde{x}_i,y_i)\}_{i=1}^{\gauss{\frac{s}{2}}}}$ \hfill \tcp{$\OIP{\tfrac{s}{2}}$} \label{line:IP_IP1}
$\IP{2} \gets \AIP{\{(\tilde{x}_i,y_i)\}_{i=\gauss{\frac{s}{2}}+1}^{s}}$ \hfill \tcp{$\OIP{\tfrac{s}{2}}$} \label{line:IP_IP2}
\Return{$\IP{1} \mul \MSP{\langle B \rangle} + \IP{2} \mul \MSP{\langle A \rangle}$} \hfill \tcp{$2 \cdot \OMulSkew{\tfrac{s}{2}}$} \label{line:IP_mul}
}
\caption{$\AIP{\{(x_i,y_i)\}_{i=1}^{s}}$}
\label{alg:IP}
}

\subsection{Concluding Remarks}
In this section, we have presented fast algorithms for the division, $q$-transform, MSP, MPE and interpolation with subquadratic complexity in $s$ over $\Fqm$, independent of $m$ (cf.~Table~\ref{tab:overview_operations_linearized_new} on page~\pageref{tab:overview_operations_linearized_new}). Our fast algorithms are faster than all previously known algorithms when $s \leq m$ (see previous work in Section~\ref{subsec:previous_work}).

After the initial submission of this paper, the preprint \citep{caruso2017fast} proposed a fast algorithm for multiplication of skew polynomials of complexity $\BigOtext{s^{\omega-2}m^2}$ over $\Fq$, which improves upon the multiplication algorithm in Section~\ref{subsec:division} for $m^{2/(5-\omega)} \leq s \leq m$. As an immediate consequence, also the cost bound for division is improved to $\BigOTtext{s^{\omega-2}m^2}$ over $\Fq$ in this range.
However, the result does not improve our cost bounds for the $q$-transform, MSP, MPE, and interpolation since the first is already quasi-linear and the other algorithms' complexity would be dominated by the $s^{\log_2(3)}$ factor, cf.~Table~\ref{tab:overview_operations_linearized_new} on page~\pageref{tab:overview_operations_linearized_new}.

\section{An Optimal Multiplication Algorithm for $s=m$}
\label{sec:implications_and_optimality}

Since the algorithms in Section~\ref{sec:fast_algos} rely on fast multiplication of linearized polynomials, we would like to know a lower bound on the cost of it. In this section, we therefore show that $m \times m$ matrix multiplication can be reduced to multiplication of linearized polynomial of degree at most $s=m$ if an elliptic normal basis of $\Fqm$ exists. This gives a lower bound on the cost of solving the latter problem.

We also speed up the algorithm for linearized polynomial multiplication modulo $x^{[m]}-x$ from \cite[Section~3.1.3]{wachter2013decoding} and show that it achieves this optimal complexity for $s=m$ and is faster than the fragmentation-based multiplication algorithm from \cite[Section~3.1.2]{wachter2013decoding} for\footnote{For $s \geq m/2$, the algorithms' outputs might differ due to the modulo operation, so they cannot be compared} $m^{2 \frac{\omega-1}{\omega+1}} \leq s < m/2$.
As a by-product, we show that the MPE at a basis of $\Fqm$ from \cite[Section~3.1.3]{wachter2013decoding} can be implemented in sub-cubic time using our fast $q$-transform algorithm from Section~\ref{subsec:fast_qtrans}.

\subsection{Relation of Linearized Polynomial Multiplication and Matrix Multiplication}

As a first step, we summarize and slightly reformulate the statements of \citep[Section~3.1.3]{wachter2013decoding} which imply that matrix multiplication and linearized polynomial multiplication are closely connected.

\begin{lemma}\label{lem:evaluation-maps-diff}
	The evaluation maps $\ev_a, \ev_b : \Fqm \to \Fqm$ of $a,b \in \Lset$ are the same if and only if $a \equiv b \mod (x^{[m]}-x)$.
\end{lemma}

\begin{pf}
Let $\mathcal{B} = \{\beta_1,\dots,\beta_m\}$, be an $\Fq$-basis of $\Fqm$ and suppose $\ev_a =\ev_b$. Then, the remainder of $a-b$ right-divided by $x^{[m]}-x$ must be zero because it vanishes on the basis $B$ and has degree smaller than $m$. The other direction is clear due to $\ev_{x^{[m]}-x} = 0$.
\end{pf}

Lemma~\ref{lem:evaluation-maps-diff} implies that the evaluation map provides a bijection between $\Lsetsmallerm$ and the set $\End_{\Fq}(\Fqm)$ of $\Fq$-linear maps $\Fqm \to \Fqm$.
Furthermore, the multiplication modulo $x^{[m]}-x$, denoted by $\mulmodm$ of two polynomials $a,b \in \Lsetsmallerm$ corresponds to the composition $\circ$ of their evaluation maps since since $\ev_{a \cdot b} = \ev_{a} \circ \ev_b$.
Using the matrix representation $[\psi]_\mathcal{B}^\mathcal{B}$ of a linear map $\psi \in \End_{\Fq}(\Fqm)$, we obtain a monoid isomorphism
\begin{align*}
\varphi_\mathcal{B} : \left( \Lsetsmallerm, \mulmodm \right) \to \left(\Fq^{m \times m}, \, \cdot \, \right), \; a \mapsto \left[\ev_a\right]_\mathcal{B}^\mathcal{B},
\end{align*}

Thus, multiplication of matrices in $\Fq^{m\times m}$ is equivalent to multiplication modulo $x^{[m]}-x$ in $\Lsetsmallerm$ and either operation can be efficiently reduced to the other, given that $\varphi_\mathcal{B}$ and its inverse can be computed fast.

Note that $\varphi_\mathcal{B}(a)$ can be computed by evaluating $a$ at the elements of $\mathcal{B}$ and representing the result in the basis $\mathcal{B}$.
The inverse mapping $\varphi_\mathcal{B}^{-1}(\ve A)$ corresponds to finding the polynomial that evaluates to the values represented by the columns of the matrix $\ve A$ (in the basis $\mathcal{B}$) at the elements of $\mathcal{B}$, i.e., an interpolation. 
Both maps can be efficiently computed as follows.
\begin{lemma}\label{lem:varphi_B_complexity}
Let $\B$ be a basis of $\Fqm$ over $\Fq$, $a \in \Lsetsmallerm$, and $\A \in \Fq^{m \times m}$.
\begin{itemize}
\item If $\B$ is a normal basis, then $\varphi_\B(a)$ (or $\varphi_\B^{-1}(A)$) can be computed by a $q$-transform (or an inverse $q$-transform).
\item Otherwise, $\varphi_\B(a)$ (or $\varphi_\B^{-1}(A)$) can be computed by a $q$-transform (or an inverse $q$-transform), plus two matrix multiplications.
\end{itemize}
\end{lemma}

\begin{pf}
Recall that $\varphi_\B(a)$ can be obtained by a multi-point evaluation at the basis $\B$ and by representing the result in the basis $\B$.
If $\B$ is a normal basis, this corresponds to a $q$-transform.
In the other cases, we can choose a normal basis $\B'$ of $\Fqm$ over $\Fq$ and first compute $\varphi_{\B'}(a)$.
Then, we use two matrix multiplications by the change of basis matrices $\T_{\B'}^{\B}$ (from $\B'$ to $\B$) and $\T_{\B}^{\B'}$ to obtain
\begin{align*}
\varphi_\B(a) = \left[ \ev_a \right]_\B^\B = \T_{\B'}^{\B} \cdot \left[ \ev_a \right]_{\B'}^{\B'} \cdot \T_{\B}^{\B'} = \T_{\B'}^{\B} \cdot \varphi_{\B'}(a) \cdot \T_{\B}^{\B'}.
\end{align*}
Analogously, we can compute $\varphi_\B^{-1}(A)$ by two matrix multiplications and an inverse $q$-transform instead of an interpolation.
\end{pf}

Lemma~\ref{lem:varphi_B_complexity} implies that any MPE and interpolation w.r.t.\ a basis of $\Fqm$ can be computed in $\BigOTtext{m^\omega}$ operations over $\Fq$.
In the following two subsections, we use this observation to speed up the multiplication algorithm modulo $x^{[m]}-x$ from \cite{wachter2013decoding} and show that it has optimal complexity.

\subsection{Faster Implementation of an Existing Multiplication Algorithm}
\label{subsec:speed_up_existing}

The following theorem shows how to speed up the algorithm for linearized polynomial multiplication modulo $x^{[m]}-x$ from \cite[Section~3.1.3]{wachter2013decoding} using our fast $q$-transform algorithm from Theorem~\ref{thm:qt-compl}.
The resulting complexity bottleneck is thus a matrix multiplication instead of a $q$-transform.

\begin{theorem}\label{thm:implication_on_multiplication}
Using the $q$-transform as described in Theorem~\ref{thm:qt-compl}, multiplication of $a,b \in \Lsetsmallerm$ modulo $x^{[m]}-x$ can be implemented in $\BigOtext{m^\omega}$ operations over $\Fq$.
\end{theorem}

\begin{pf}
By the properties of $\varphi_\B$, we can compute $c = a \cdot b \mod (x^{[m]}-x)$ by
\begin{align*}
c = \varphi_\B^{-1} \left( \varphi_\B(a) \cdot \varphi_\B(b) \right).
\end{align*}
The computations of the two $\varphi_\B$ and one $\varphi_\B^{-1}$ cost six matrix multiplications, a $q$-transform and an inverse $q$-transform.
In addition, we need to perform a matrix multiplication.
Using the algorithm for $q$-transform described in Theorem~\ref{thm:qt-compl} together with the bases from \citep{couveignes2009elliptic}, the (inverse) $q$-transform costs $\BigOTtext{m^2}$ operations over $\Fq$. Hence, the matrix multiplications with complexity $\BigOtext{m^\omega}$ over $\Fq$ are dominant.
\end{pf}

\begin{remark}
For polynomials of $q$-degree $s<m/2$, the algorithm described above is a linearized polynomial multiplication algorithm since the result has degree $<m$ and is not affected by the modulo $x^{[m]}-x$ reduction.
It is possible to extend the algorithm to polynomials of $q$-degree $s\geq m/2$ as follows.

Let $s = \mu \cdot m/2$ for $\mu\geq1$. Then, we can fragment $a,b$ into $\mu$ polynomials of degree $<m/2$. We then pairwise multiply the fragments of $a$ and $b$ respectively (costs $\mu^2$ many multiplications of degree $<m/2$ polynomials: In total, $\BigOtext{\mu^2m^\omega}$). Addition of the fragments is negligible since we know the overlapping positions. Hence, we obtain a complexity of
\begin{align*}
\BigO{\max\left\{s^2m^{\omega-2},m^\omega\right\}}
\end{align*}
in operations over $\Fq$. For $m^{2\frac{\omega-1}{\omega+1}} < s < m^2$, this multiplication algorithm is faster than the one of \citep[Section~3.1.2]{wachter2013decoding} (see Section~\ref{subsec:division}), which has complexity $\BigOTtext{m s^{\polymulexponent}}$ over $\Fq$ when using the bases of \citet{couveignes2009elliptic}.
In addition, the constant hidden by the $\mathcal{O}$-notation is smaller since the matrix multiplication is with respect to $m$, which is much larger than $\sqrt{s}$ in the case of the algorithm in Section~\ref{subsec:division} (cf.~Remark~\ref{rem:pracical_complexities}) for $s \approx m$.
\end{remark}

\subsection{Optimal Multiplication Algorithm for $s=m$}

We prove the optimality of the algorithm described in Theorem~\ref{thm:implication_on_multiplication} by reducing matrix multiplication to polynomial multiplication.
Lemma~\ref{lem:varphi_B_complexity} implies that if the basis in which we represent elements of $\Fqm$ is a normal basis, we can reduce matrix multiplication to a $q$-transform, two inverse $q$-transforms and a multiplication of two linearized polynomials modulo $x^{[m]}-x$ (note that the modulo reduction only requires Frobenius automorphisms and $\BigOtext{m}$ many additions in $\Fqm$, so in total $\BigOtext{m^2}$ operations in $\Fq$).

In addition, when the bases admit quasi-linear multiplication as the so-called normal elliptic bases from \citep{couveignes2009elliptic}, the $q$-transform only costs $\BigOTtext{m^2}$ operations over $\Fq$ by Theorem~\ref{thm:qt-compl}, and the complexity bottleneck becomes the multiplication of two linearized polynomials of degree $m$. This can be summarized in the following statement.
Let $\OMul{m}$ denote the worst-case cost of multiplying two polynomials from $\Lsetsmallerm$ (note that the polynomial degrees are smaller than $s=m$).

\begin{lemma}\label{lem:matrix_mult_red_poly_mult}
Let $q,m$ be such that there is an elliptic normal basis of $\Fqm$ over $\Fq$. Then, the multiplication of two matrices from $\Fq^{m \times m}$ can be implemented in
\begin{align*}
\BigO{m^\omega} \subseteq \BigO{\OMul{m}}
\end{align*}
operations over $\Fq$.
\end{lemma}

Lemma~\ref{lem:matrix_mult_red_poly_mult} states that if a normal elliptic basis exists for $q,m$, then matrix multiplication can be efficiently reduced to linearized polynomial multiplication in $\Lsetsmallerm$.
Such bases do not exist for all pairs $(q,m)$.
However, we can give the following statement.

\begin{lemma}\label{lem:m_q_sequence}
Let $(m_i)_{i \in \NN}$ be a sequence of $m_i \in \NN$ with $m_i \to \infty$ $(i \to \infty)$. Then, there is a sequence $(q_i)_{i \in \NN}$, where the $q_i$ are prime powers, such that there is an elliptic normal basis of $\mathbb{F}_{q_i^{m_i}}$ over $\mathbb{F}_{q_i}$.
\end{lemma}

\begin{pf}
Let $\tilde{q}_i$ be any sequence of prime powers. Due to \citep[Section~5.2]{couveignes2009elliptic}, there is a positive integer $f_i \in \BigOtext{\log^2(m_i)(\log\log(m_i))^2}$ such that $q_i = \tilde{q}_i^{f_i}$ admits an elliptic normal basis as desired.
\end{pf}

Suppose that $\OMul{m} \in \Theta(m^\gamma)$ for some $\gamma\geq 2$, independent of the ground field $q$.
Let $(q_i,m_i)_{i \in \NN}$ be a sequence of pairs $q_i$ and $m_i$ as in Lemma~\ref{lem:m_q_sequence}.
Then, by Lemma~\ref{lem:matrix_mult_red_poly_mult} there must be a constant $C \in \mathbb{R}_{>0}$ and an index $j \in \NN$ such that
\begin{align*}
m_i^\omega \leq C \cdot m_i^\gamma
\end{align*}
for all $i\geq j$. Hence, $\omega \leq \gamma$, which provides a lower bound for the linearized polynomial multiplication exponent $\gamma$.
Note that the multiplication algorithm in Section~\ref{subsec:speed_up_existing} achieves this bound and therefore has optimal complexity.

The fragmentation algorithm described in \citep{wachter2013decoding} (see also Section~\ref{subsec:division}), in combination with the bases of \citep{couveignes2009elliptic} achieves $\gamma=\tfrac{\omega+3}{2}$, so its complexity differs from an optimal solution by a factor $m^{\frac{3-\omega}{2}}$ (note that this only holds for $s=m$).

Note that our argumentation also implies that the existence of a linearized polynomial multiplication algorithm with quasi-linear complexity in $s$ over $\Fqm$, independent of $s$, would give a quasi-quadratic matrix multiplication algorithm in the cases where an elliptic normal basis of $\Fqm$ exists.
Hence, proving that a quasi-linear linearized polynomial multiplication algorithm exists is at least as hard as proving that matrix multiplication can be implemented in quasi-quadratic time.

\section{Decoding Gabidulin Codes in Sub-quadratic Time}\label{sec:decoding}
Gabidulin codes are rank-metric codes that can be found in a wide range of applications, including network coding~\citep{silva2008rank}, code-based cryptosystems~\citep{gabidulin1991ideals}, and distributed storage systems~\citep{SilbersteinRawatVish-ErrorResilDistributedStorage_2012}.

In this section, we show that two algorithms for decoding Gabidulin codes from \citep{wachter2013decoding}, one for only errors and one including generalized row and column erasures, can be implemented in $\BigOTtext{n^{\max\left\{\log_2(3),\polymulexponent\right\}}}$ operations over $\Fqm$ using the methods presented in this paper.
This yields the first decoding algorithms for Gabidulin codes with sub-quadratic complexity.

\subsection{Notation}

A rank-metric code $\mycode{C} \subseteq \Fq^{m \times n}$ is a set of matrices over a finite field $\Fq$, where the distance of two codewords is measured w.r.t.\ the \emph{rank distance}
\begin{align*}
\dR : \Fq^{m \times n} \times \Fq^{m \times n} \to \NN_0, \quad (\C_1,\C_2) \mapsto \rank(\C_1-\C_2).
\end{align*}
Since, for a fixed $\Fq$-basis of $\Fqm$, elements in $\Fqm^n$ can be expanded into matrices in $\Fq^{m \times n}$, the rank distance is also well-defined over $\Fqm^n$. A linear rank-metric code of length $n$, dimension $k$, and minimum rank distance $d_R$ is a $k$-dimensional $\Fqm$-subspace of $\Fqm^n$ whose elements have pairwise rank distance at least $d_R$.
It was shown in \citep{Delsarte_1978,Gabidulin_TheoryOfCodes_1985,Roth_RankCodes_1991} that any such code with $n \leq m$ fulfills the rank-metric Singleton bound $d_R \leq n-k+1$. Codes achieving this bound with equality are called maximum rank distance (MRD) codes.

\emph{Gabidulin codes} \citep{Delsarte_1978,Gabidulin_TheoryOfCodes_1985,Roth_RankCodes_1991} are a special class of MRD codes and are often considered as the analogs of Reed--Solomon codes in rank metric.
They can be defined by the evaluation of degree-restricted linearized polynomials as follows.
\begin{definition}[\citet{Gabidulin_TheoryOfCodes_1985}]\label{def:gabidulin_code}
	A linear Gabidulin code $\Gab{n,k}$ over $\Fqm$ of length $n\leq m$ and dimension $k \leq n$ is the set
	\begin{equation*}
	\Gab{n,k} = \Big\lbrace \; \big[\begin{matrix}
	f(g_1) & f(g_2) & \dots & f(g_{n})
	\end{matrix}\big] : f \in \Lsmallerk \;\Big\rbrace,
	\end{equation*}
	where the fixed elements $g_1,g_2, \dots, g_{n} \in \Fqm$ are linearly independent over $\Fq$. 
\end{definition}

Note that the \emph{encoding} of Gabidulin codes, see \cref{def:gabidulin_code}, is equivalent to the calculation of one MPE and can therefore be accomplished with complexity $\BigOtext{s^{\polymulexponent} \log(s)}$. If $\{g_1,\dots,g_n\}$ is a normal basis, it can be computed as a $q$-transform in $\BigOtext{s \log^2(s) \log(\log(s))}$.

In this section, we assume that a word $\r = \c+\e$ is received, where $\dR(\r,\c) = \rk(\e)$ denotes the number of rank errors, and the decoder wants to retrieve $\c$ from $\r$.

\subsection{Error-Only Decoding of Gabidulin Codes}
Algorithm~\ref{alg:gaoalgo} shows \citep[Algorithm~3.6]{wachter2013decoding} for decoding Gabidulin codes up to $\left\lfloor(d-1)/2\right\rfloor$ rank errors.
This algorithm can be seen as the rank-metric equivalent of the Reed--Solomon decoding algorithms from \citep{Sugiyama_AMethodOfSolving_1975,welch_error_1986,Sudan:JOC1997,Gao_ANewDecodingAlgorithm_2002}.
Its correctness was proven in \citep[Theorem~3.7]{wachter2013decoding} and its complexity was shown to be in $\BigOtext{n^2}$ over $\Fqm$.
Since all steps of Algorithm~\ref{alg:gaoalgo} can be performed by algorithms with sub-quadratic complexity from this paper, the following corollary holds.

\begin{corollary}
Algorithm~\ref{alg:gaoalgo} can be implemented in
$\BigO{n^{\max\left\{\log_2(3),\polymulexponent\right\}}\log^2(n)}$ operations over $\Fqm$.
\end{corollary}

\printalgoIEEE{
\DontPrintSemicolon
\KwIn{Received word $\ve{r} \in \Fqm^n$ and $g_1,g_2,\dots,g_n\in \Fqm$, linearly independent over $\Fq$.}
\KwOut{Estimated evaluation polynomial $f$ with $\qdeg f < k$ and error span polynomial $\Lambda$ or ``decoding failure''.}
$\qtr{r} \leftarrow \IP{\{(g_i,r_i)\}_{i=1}^{n}}$ \hfill \tcp{$\OIP{n}$} \label{line:interpol}
$\mathcal{M} \gets \MSP{\LH{g_1,\dots,g_n}}$ \hfill \tcp{$\OMSP{n}$} \label{line:leea-msp}
$[\LEEAOutputRx, \LEEAOutputUx, \LEEAOutputVx]$ $\leftarrow$ \textsc{RightLEEA}$\big(\mathcal{M},  \qtr{r}, \npluskhalf\big)$ \hfill \tcp{$\OMulSkew{n} \log^2(n)$ (Corollary~\ref{cor:compl-leea})} \label{line:leea}
$[\qL, \rL] \leftarrow$ Left-divide $\LEEAOutputRx$ by $\LEEAOutputUx$ \hfill \tcp{$\ODiv{n}$} \label{line:left-div}
\If{$\rL = 0$}{\Return{$[f,\Lambda] \gets [\qL,\LEEAOutputUx]$}\;}
\Else{
\Return{``decoding failure''}\;
}
\caption{$\mathrm{DecodeGabidulin}\big(\vec{r}, \{g_1,g_2,\dots,g_n \}\big)$}
\label{alg:gaoalgo}
}

\subsection{Error-Erasure Decoding}

Algorithm~\ref{alg:gaoalgo_ee} shows \citep[Algorithm~3.7]{wachter2013decoding} for decoding Gabidulin codes with $t$ errors, $\rho$ generalized row erasures and $\gamma$ generalized column erasures if
\begin{equation*}
2t+\rho+\gamma \leq d-1,
\end{equation*}
where $d$ is the minimum rank distance of the Gabidulin code.
The correctness of Algorithm~\ref{alg:gaoalgo_ee} was proven in \citep[Theorem~3.9]{wachter2013decoding}.

\begin{theorem}
Algorithm~\ref{alg:gaoalgo_ee} can be implemented in $\BigO{n^{\max\left\{\log_2(3),\polymulexponent\right\}}\log^2(n)}$ operations over $\Fqm$.
\end{theorem}

\begin{pf}
Its lines have the following complexities:
\begin{itemize}
\item Line~\ref{line:gao_ee_a}: $d^{(C)}_i \in \Fqm$ and if $\Fqm$ elements are represented in the normal basis generated by $\Normelement$, the $B^{(C)}_{i,j}$'s are already the representation of $d^{(C)}_i$, and thus no computation is needed.
\item Lines~\ref{line:gao_ee_b} and~\ref{line:gao_ee_c} calculate MSPs whose cost is in $\OMSP{n} \subseteq \BigOtext{n^{\max\left\{\log_2(3),\polymulexponent\right\}} \log(n)}$.
\item The cost of Line~\ref{line:gao_ee_d} is negligible.
\item Line~\ref{line:gao_ee_e} finds the interpolation polynomial of $n$ point tuples, implying a cost of $\OIP{n} \subseteq \BigOtext{n^{\max\left\{\log_2(3),\polymulexponent\right\}} \log(n)}$. 
\item Line~\ref{line:gao_ee_f} requires three multiplications of linearized polynomials of degree $\leq 3n$ plus the modulo operation which requires $\BigO{m} \subseteq \BigO{n}$ additions because $x^{[m]}-x$ has only two non-zero coefficients. Hence, its complexity lies in $\BigOtext{\OMulSkew{n}} \subseteq \BigOtext{n^{\polymulexponent}}$.
\item Line~\ref{line:gao_ee_g} has complexity $\BigOtext{n^{\max\left\{\log_2(3),\polymulexponent\right\}} \log^2(n)}$ by Corollary~\ref{cor:compl-leea}.
\item Lines~\ref{line:gao_ee_h} and~\ref{line:gao_ee_i} compute a multiplication of polynomials of degree $\leq n$ and a left division, yielding a complexity of $\BigOtext{\ODiv{n}} \subseteq \BigOtext{n^{\polymulexponent} \log(n)}$.
\end{itemize}
Thus, using the results of \cref{sec:fast_algos}, the overall complexity is as stated.
\end{pf}

\printalgoIEEE{
\DontPrintSemicolon
\KwIn{Received word $\vec{r} \in \Fqm^n$,\newline
$g_i = \beta^{[i-1]} \in \Fqm$, $i=1,\dots,n$, normal basis of $\Fqm$ over $\Fq$,\newline
$\vec{a}^{(R)} = \vecelementsArb{a^{(R)}}{\numbRowErasures} \in \Fqm^\numbRowErasures$;\newline
$\Mat{B}^{(C)} = \big[B_{i,j}\big]^{i\in \intervallincl{1}{\numbColErasures}}_{j\in \intervallincl{1}{n}} \in \Fq^{\numbColErasures \times n}$
}
\KwOut{Estimated evaluation polynomial $f$ with $\qdeg f < k$ or ``decoding failure''.}
$d^{(C)}_i\leftarrow\sum_{j=1}^{n}B^{(C)}_{i,j}\Normelement^{[j-1]}$ for all $i=1,\dots,\numbColErasures$ \hfill \tcp{negligible} \label{line:gao_ee_a}
$\Gamma^{(C)} \leftarrow \MSP{\LH{d^{(C)}_1,d^{(C)}_2, \dots,d^{(C)}_{\numbColErasures}}}$ \hfill \tcp{$\OMSP{\numbColErasures} \subseteq \OMSP{n}$} \label{line:gao_ee_b}
$\Lambda^{(R)} \leftarrow \MSP{\LH{a^{(R)}_1,a^{(R)}_2, \dots,a^{(R)}_{\numbRowErasures}}}$ \hfill \tcp{$\OMSP{\numbRowErasures} \subseteq \OMSP{n}$} \label{line:gao_ee_c}
$\qreciproc{\Gamma^{(C)}} \gets \sum_{i=0}^{m-1}\qreciproc{\Gamma^{(C)}_i} x^{[i]}$ with $\qreciproc{\Gamma^{(C)}_i} := \Gamma_{-i \mod m}^{(C)[i]}$ for all $i = 0,\dots,m$ \hfill \tcp{$\BigO{m} \subseteq \BigO{n}$}  \label{line:gao_ee_d}
$\qtr{r} \leftarrow \IP{\{(g_i,r_i)\}_{i=1}^{n}}$ \hfill \tcp{$\OIP{n}$} \label{line:gao_ee_e}
$\qtr{y} \gets \Lambda^{(R)} \cdot \qtr{r} \cdot \qreciproc{\Gamma^{(C)}} \cdot x^{[\numbColErasures]} \mod (x^{[m]}-x)$ \hfill \tcp{$\ODiv{n}$}  \label{line:gao_ee_f}
$[\LEEAOutputRx,\LEEAOutputUx,\LEEAOutputVx] \gets \ALEEA{x^{[m]}-x, \qtr{y}, \left\lfloor\frac{n+k+\numbRowErasures+\numbColErasures}{2}\right\rfloor }$ \hfill \tcp{$\OMulSkew{n} \log^2(n)$}  \label{line:gao_ee_g}
$[\qL, \rL] \gets \ALDIV{\LEEAOutputRx, \LEEAOutputU \cdot \Lambda^{(R)}}$ \hfill \tcp{$\ODiv{n}$}  \label{line:gao_ee_h}
$[\qR, \rR] \gets \ARDIV{\qL, \qreciproc{\Gamma^{(C)}} \cdot x^{[\numbColErasures]} \mod (x^{[m]}-x)}$ \hfill \tcp{$\ODiv{n}$}  \label{line:gao_ee_i}
\If{$\rL = 0$ \textup{\textbf{and}} $\rR = 0$}{\Return{$f \gets \qL$}\;}
\Else{\Return{``decoding failure''}\;}
\caption{$\mathrm{DecodeErrorErasureGabidulin}\big(\vec{r}, \{g_1,g_2,\dots,g_n\}, \vec{a}^{(R)}, \Mat{B}^{(C)} \big)$}
\label{alg:gaoalgo_ee}
}

\begin{remark}
In both Algorithm~\ref{alg:gaoalgo} and \ref{alg:gaoalgo_ee}, the involved polynomials have $q$-degree at most $n \leq m$. Hence, the new algorithms in this paper are asymptotically faster than the ones from \citep{caruso2017new} in this case.
\end{remark}

\section{Conclusion}\label{sec:conclusion}

In this paper, we have reduced the complexity of several operations with linearized polynomials. 
Table~\ref{tab:overview_operations_linearized_new} on page~\pageref{tab:overview_operations_linearized_new} summarizes the new complexity bounds on operations with linearized polynomials.
In particular, we have generalized a fast algorithm for linearized polynomial multiplication to skew polynomials, implying the first sub-quadratic algorithm for division of two linearized polynomials of degree $s\leq m$.
We have also presented new algorithms with sub-quadratic complexity for calculating the $q$-transform, minimal subspace polynomial computation, multi-point evaluation and interpolation.
For the case $s=m$, we have presented a lower bound on the cost of linearized polynomial multiplication and an algorithm that achieves it.
Further, we have shown how to apply these algorithms when decoding Gabidulin codes. This yields the first decoding algorithm of Gabidulin codes which has, over $\Fqm$, sub-quadratic complexity in the code-length.

For future work, it is interesting to include our new algorithms in the study from~\citep{BohaczukSilva-EvaluationErasureDecodingGabidulin} on fast erasure decoding of Gabidulin codes.

\section*{Acknowledgement}
The authors would like to thank Johan Rosenkilde n\'e Nielsen for the valuable discussions, his idea leading to \cref{alg:IP} and several comments improving the readability of the paper.
Also, we would like to thank Luca~De~Feo for making us aware of \citep{couveignes2009elliptic}.

The authors also want to emphasize that the reviewers' comments were extremely helpful and thoughtful, and want to thank them for their excellent comments. To give one example, one of the reviewer's suggestions significantly shortened the proof of Theorem~\ref{thm:mspmpe-compl}.

Sven Puchinger's work was supported by the German Research Foundation (Deutsche Forschungsgemeinschaft, DFG), grant BO~867/29-3.
This work was partly done while Antonia Wachter-Zeh was with the Technion---Israel Institute of Technology.
At the Technion, Antonia Wachter-Zeh's work was supported by the European Union’s Horizon 2020 research and innovation programme under the Marie Sklodowska-Curie grant agreement No. 655109.
At TUM, Antonia Wachter-Zeh's work was supported by the Technical University of Munich---Institute for Advanced Study, funded by the German Excellence Initiative and European Union Seventh Framework Programme under Grant Agreement No.~291763 and an Emmy Noether research grant from the German Research Foundation (Deutsche Forschungsgemeinschaft, DFG), grant WA~3907/1-1.

\bibliographystyle{elsarticle-harv}
\bibliography{bib}

\end{document}